\documentstyle[11pt,aaspp4]{article}

\slugcomment{Accepted for A.J. 3 May 02}
\lefthead{Dahn et al.}
\righthead{Cool Dwarfs and Brown Dwarfs}

\begin{document}

\title{Astrometry and Photometry for Cool Dwarfs and Brown Dwarfs}

\author{Conard C. Dahn, Hugh C. Harris, Frederick J. Vrba,
Harry H. Guetter, Blaise Canzian, 
Arne A. Henden\footnote{Universities Space Research Association.},
Stephen E. Levine, Christian B. Luginbuhl, Alice K. B.  Monet,
David G. Monet, Jeffrey R. Pier, Ronald C. Stone, 
and Richard L. Walker}
\affil{U.S. Naval Observatory, PO Box 1149, Flagstaff, AZ 86002-1149}

\author{Adam J. Burgasser\footnote{
Hubble Postdoctoral Research Fellow.}}
\affil{Dept. of Astronomy and Astrophysics, UCLA, 8965 Math Sciences
Bldg., Los Angeles, CA 90095-1562}

\author{John E. Gizis\footnote{
NOAO WIYN Queue Investigator, Kitt Peak National Observatory,
National Optical Astronomy Observatories.  NOAO is operated by the
Association of Universities for Research in Astronomy, Inc.,
under cooperative agreement with the National Science Foundation.}}
\affil{Dept. of Physics and Astronomy, University of Delaware,
Newark, DE 19716}

\author{J. Davy Kirkpatrick}
\affil{Infrared Processing and Analysis Center, MS 100-22,
Pasadena, CA 91125}

\author{James Liebert}
\affil{Steward Observatory, University of Arizona, 933 North Cherry
Avenue, Tucson, AZ 85721}

\author{I. Neill Reid}
\affil{Space Telescope Science Institute, 3700 San Martin Drive,
Baltimore, MD 21218}

\begin{abstract}
Trigonometric parallax determinations are presented for 28 late type dwarfs
and brown dwarfs, including eight M dwarfs with spectral types between M7
and M9.5, 17 L dwarfs with spectral types between L0 and L8, and three T
dwarfs.  Broadband photometry at CCD wavelengths ($VRIz^{*}$) and/or near-IR
wavelengths ($JHK$) are presented for these objects and for 24 additional
late-type dwarfs.  Supplemented with astrometry and photometry from the
literature, including ten L and two T dwarfs with parallaxes established
by association with bright, usually HIPPARCOS primaries,
this material forms the basis for studying various color$-$color and
color$-$ absolute magnitude relations.
The $I-J$ color is a good predictor of absolute magnitude for late-M and
L dwarfs.  $M_J$ becomes monotonically fainter with $I-J$ color and with
spectral type through late-L dwarfs, then brightens for early-T dwarfs.
The combination of $z^*JK$ colors alone can be used to classify
late-M, early-L, and T dwarfs accurately, and to predict their absolute
magnitudes, but is less effective at untangling the scatter among mid- and
late-L dwarfs.  The mean tangential velocity of these objects is found
to be slightly less than that for dM stars in the solar neighborhood,
consistent with a sample with a mean age of several Gyr.
Using colors to estimate bolometric corrections, and models to estimate
stellar radii, effective temperatures are derived.  The latest L dwarfs
are found to have $T_{\rm eff} \sim 1360$~K.

\end{abstract}


\section{INTRODUCTION}

The first identification of a low-mass dwarf cooler than the well-know
M dwarfs came with the discovery of a very red companion to the white
dwarf GD165 by Becklin and Zuckerman (1988).  Initial spectroscopic
observations at far-red wavelengths ($\lambda\lambda\,$6300$-$9000 \AA)
were puzzling because they showed none of the features characteristic
of very late M dwarfs (Kirkpatrick et al. 1993).  Subsequently, a faint
companion to the bright M1V star Gliese~229 was discovered
(Nakajima et al.\ 1995), and its near-IR spectrum was found to be
dominated by deep absorption bands of methane (Oppenheimer et al. 1995),
more similar to the planet Jupiter than to the known late-type, low-mass
stars.  The low luminosity of both objects could be measured through
trigonometric parallaxes of their bright primary companions.  Their
spectra became prototypes for two new spectral classes, now designated
L (having strong lines of neutral alkali elements and hydrides) and
T (having strong bands of methane at near-IR wavelengths, and sometimes
called ``methane dwarfs''), both indicating cooler effective temperatures
than type M (having strong molecular bands of TiO and VO).
Several excellent papers review the rapid development of the
observational data (Reid \& Hawley 2000; Basri 2000; Liebert 2000)
and theoretical understanding
(Hauschildt et al. 1997; Chabrier \& Baraffe 2000; Burrows et al. 2001)
of these low-mass objects that bridge the gap between
cool M stars and giant Jupiter-like planets.

Rapid growth of this topic has been spurred by the identification of
{\it many} new L and T dwarfs from ongoing survey programs DENIS (Epchtein
et al.\ 1997), 2MASS (Skrutskie et al.\ 1997), and SDSS (York et al. 2000).
Initial results from DENIS (Delfosse et al.\ 1997) and 2MASS (Kirkpatrick
et al. 1997a) revealed several objects which follow-up spectroscopy showed
ranged from comparable to GD165B (DEN1058$-$15) to definitely
cooler than GD165B, but definitely warmer than Gl~229B (DEN0205$-$11).
By early 2000, the number of recognized L dwarfs had rapidly grown to nearly
100 (Kirkpatrick et al. 2000; hereafter, K00).  However, the identification
of additional T dwarfs like Gl~229B proceeded much more slowly,
despite considerable efforts searching for them.  Finally, in mid-1999,
several were identified, two by the SDSS collaboration
(Strauss et al.\ 1999; Tsvetanov et al.\ 2000; Pier et al.\ 2000) and
four by the 2MASS collaboration (Burgasser et al. 1999; Burgasser 2000c).
As of this writing, a total of 30 T dwarfs have been found
(Burgasser et al.\ 2002; a few not yet published).

As objects like GD165B became known, the logical first step
in understanding this new population was to classify them into a spectral
sequence.  Kirkpatrick (1998) discussed at some length the need for
one or more new spectral classes to accomodate objects cooler than the
M dwarfs.  He included ``L'' as a possible appropriate designation
(see also Mart\'{\i}n et al. 1997).
Detailed classification systems were then developed independently by
Mart\'{\i}n et al.\ (1999c) and Kirkpatrick et al.\ (1999; hereafter, K99).
The system proposed by Mart\'{\i}n and coworkers is based on pseudocontinuum
flux ratios in the red (F$_{\rm \lambda 825nm}$/F$_{\rm \lambda 756nm}$)
calibrated to the temperature scale derived by Basri et al. (2000).
Kirkpatrick's system maintains the spirit of the long standing MK
classification scheme and is strictly morphological, relying
on the progressive strengthening and weakening of various spectral features
and ratios within the $\lambda\lambda$6300--10000 \AA\ window from
$\sim\,$10 \AA\ resolution spectra.
The two scales agree on average for earlier types through about L2V
(although they often disagree for individual objects), but diverge
for later types, with L7.5V on the Kirkpatrick scale corresponding roughly
to L6V on the Mart\'{\i}n et al. scale.

Similarly, the Gl~229B-like objects have been grouped together
into a spectral class designated as ``T,'' following the proposal
by Kirkpatrick (K99).  Classification by the 2MASS group (Burgasser
et al.\ 2002) uses indices measuring
the strengths of the CH$_{\rm 4}$ bands at 1.3 and 1.6 $\mu$m,
the H$_{\rm 2}$O/CH$_{\rm}$ band blend at 1.15 $\mu$m, along with the flux
peak ratios between the $zJK$ bands, from lower resolution (R$\,\sim\,$100
at 1 $\mu$m) spectra.  Classification by the SDSS collaborators
(Geballe et al.\ 2002) employs higher resolution
spectra (R$\,\geq\,400$), measuring indices sensitive to the H$_{\rm 2}$O
and CH$_{\rm 4}$ bands at $\lambda\lambda\,$1.15-2.2 $\mu$m,
but the results are in excellent agreement with those from the 2MASS
group (with a maximum difference of only one-half spectral subclass
for 11 stars in common to the two studies).
Efforts to classify L dwarfs and the L/T transition objects from
near-IR spectra have been made by several groups (Reid et al.\ 2001;
Testi et al.\ 2001; Geballe et al.\ 2002; McLean et al.\ 2002).
Results from these studies made at different spectral resolution
and their correlation with classification made from optical spectra
are not yet understood completely.

With several classification schemes defined, the present challenge is
to better understand the atmospheric physics responsible for the various
spectral features.  Evaluating the spectral sequences in terms of
$T_{\rm eff}$ is perhaps the most basic.
A rough scale (K99) for the late-M through L dwarfs based on the observed
appearance/disappearance of various far-red ($\lambda\lambda\,$0.63--1.01
$\mu$m) spectral features compared with the theoretical atmosphere models
of Burrows and Sharp (1999) runs from $T_{\rm eff} \sim\,$2000~K at L0 to
$\sim\,$1500~K at L8.  Mart\'{\i}n et al. (1999c) established their
L-dwarf spectral classification sequence to correspond to a temperature
sequence running from 2200~K
at L0 to 1600~K at L6 (or $\sim\,$L7.5 on the Kirkpatrick system),
with $T_{\rm eff}$ values primarily from the work of Basri et al. (2000).
The rapid development of model atmospheres for such cool dwarfs has aided
the interpretation of spectroscopic data, and several other studies have
made temperature estimates (e.g. Pavlenko et al.\ 2000; Leggett et al.
2001; Stephens et al.\ 2001; Schweitzer et al.\ 2001, 2002).
However, creating realistic models is challenging, particularly in the
treatment of dust.  At present, discrepancies exist between the models and
observed spectra, leaving some doubt that current model atmospheres
alone can be used to provide accurate effective temperatures.  However,
effective temperatures of cool dwarfs are closely related to their
luminosities because (with the exception of young objects,
with ages $<0.3$~Gyr)
they have only a small range of radii, set by the equation of state
of their degenerate interiors.  This alternate method for determining
their temperatures from their luminosities was first used by K00,
and is a motivation for measuring parallaxes for additional cool dwarfs.
Temperatures determined for a few late-L dwarfs from luminosities
tend to be systematically lower than those determined from a comparison
to model atmospheres (Leggett et al. 2001; Schweitzer et al. 2002).
There is presently a significant range in the various temperature
scales for late-M through late-L dwarfs (Chabrier 2002).

Distance measurements are needed to determine the luminosities of
individual objects.  More than that, several factors make it necessary to
measure an ensemble of cool dwarfs of a given type in order to acquire
their mean luminosity (and range of luminosities), to identify outliers,
and to establish observationally the luminosity of their isochrones.  
First, any measured dwarf can always be an unresolved binary, with
colors and magnitudes measured for the composite system.
Second, field dwarfs with extreme ages can have atypical luminosities
for their color, temperature, or spectral type.  
Third, very cool dwarfs may exhibit a wider range of observational
characteristics than do warmer main sequence stars due to the presence
and variation of dust opacity.  Simple models (e.g. Burrows et al.\ 2002;
Marley et al.\ 2002) predict a sensitivity of colors for different
surface gravity, metallicity, and dust precipitation, and the
near-IR colors of L dwarfs (e.g. Leggett et al.\ 2002a) show little
correlation with spectral type.
Fourth, temporal variability is also expected if
atmospheric clouds change with time.  Observations to date indicate
only low-level variability (e.g. Bailer-Jones \& Mundt 2001),
but larger amplitudes may occur for some objects.

Substellar brown dwarfs are being studied in young clusters, but in
intermediate-age or old clusters they are too faint and rare to provide
useful samples.  A few very cool dwarfs (like GD165B and Gl~229B mentioned
above) are companions of brighter stars, where the parallax of the brighter
primary star can be easily measured.  There are insufficient numbers
of these, however, to draw general trends.  Therefore, we are left with
the need to measure trigonometric parallaxes for a substantial number
of field L and T dwarfs.  In this paper, we extend previous work
(K00; Dahn et al.\ 2000) by presenting parallax measurements for eight M,
17 L, and three T dwarfs.  Together with companions to brighter
stars with parallaxes already available, there are now 25 L dwarfs
compiled in this paper, enough to establish their isochrones with some
reliability for the first time.  Only five T dwarfs have known distances,
so for them the work is just beginning.

\section{ASTROMETRY}

\subsection{Observational Procedures}

The astrometric observations reported here were carried out as part of
the Naval Observatory CCD faint-star parallax program using the 1.55~m Strand
Astrometric Reflector.  The camera (designated TEK2K and used for all fields
except LP412$-$31) employs a thinned, back-side illuminated Tektronix
2048 x 2048 CCD with 24.0 micron pixels, corresponding to 0.325 arcsec
pix$^{-1}$.  This scale gives a field of view of 11 x 11 arcmin.
The LP412$-$31 parallax observations presented here were made during
the time period 1990.8 to 1995.8 using a thinned Texas Instruments
800 x 800 CCD (TI800).  This camera system and wide-$R$
filter have been described in detail by Monet et al. (1992).

Beginning in late 1992, the entire USNO faint-star parallax program began
transitioning to the larger format TEK2K camera due to the availablity of
improved reference star frames within the larger field of view.  As a result
of experience gained with the TI800 camera, the TEK2K chip is mounted on
the 1.55~m with the columns oriented (to a high accuracy) east-west.
This orientation takes advantage of the fact that slightly higher astrometric
accuracy appears to be obtained along CCD columns versus CCD rows, and the
fact that the larger annual parallactic shift occurs in Right Ascension.
At the time that the TEK2K observations for these fields were initiated,
the reddest astrometrically flat filter available was a wide-$I$ interference
filter with passband centered at $\sim\,$8100 \AA\ and with
FWHM$\,\sim\,$1910 \AA.  Consequently, the majority of the fields have been
observed with this filter.  In response to the difficulties encountered in
attempting to observe the extremely red T dwarfs using the 1.55~m and this
$I$ filter, an astrometrically flat SDSS $z^*$ filter was procured in 1999.
This $z^*$ filter plus CCD detector gives a passband centered at roughly
$\sim\,$8800 \AA\ and with FWHM$\,\sim\,$1100 \AA.  It has been employed
for the astrometry of the three T dwarfs presented here and for the L
dwarf 2MW0825+21; the wide-$I$ filter has been employed for all of the other
TEK2K fields.

Experience carrying out parallax astrometry with TEK2K on over 200 fields
for epoch spans as long as 9 years demonstrates that relative astrometric
measurements accurate to $\sim\pm$3 mas are routinely obtained for a single,
well-exposed, image.  Although the quality of a parallax determination depends
on many factors, including the quality of the available stellar reference frame
and particularly on the observational coverage of the parallactic ellipse, a
formal standard error for the relative parallax of $\sim\pm$0.5 mas is typically
realized with TEK2K after roughly 100 observations adequately distributed over
the parallactic ellipse and spanning at least three years in epoch range.
However, many of the targets in the present sample are far from ``typical."
In fact, several of these objects (e.g., PC0025+04, 2MW0850+10, 2MW1328+21,
SDSS1624+00, and 2MW1632+19) are so faint that nearly ideal observing
conditions -- cloudless, dark skies (moonless and well away from twilight),
and with seeing better than 1.2-1.4 arcsec FWHM -- are required to secure
astrometrically useful data; and even then the exposure times required are
60--90 min in length for only relatively low S/N results.  Together with
seasonal weather patterns, these constraints result in less than ideal
coverage of the parallatic ellipse and, as we shall see below, much less
accurate parallax determinations than those obtained for brighter stars
when using TEK2K.

The observing procedures employed in the TEK2K parallax program call for
exposures to be centered within $\pm\,$15 min of meridian transit.
Strict adherence to this rule effectively eliminates the need for a
correction for differential color refraction (DCR; cf. Monet et al. 1992).
This is because astrometry for purposes of parallax determinations is highly
differential -- all that we are interested in is the shift in position as a
function of time produced by different parallax factors as the Earth orbits
the Sun.  By taking all observations close to meridian transit (at
constant hour angle and, hence, constant zenith distance) the shift in
position of the target star relative to the reference stars caused by DCR
is, to first-order, identical for all observations on any particular target.
This is especially advantageous for astrometry of targets as extremely red
as the L and T dwarfs presented here.  As a matter of routine, a DCR
correction has been applied to the L dwarfs observed with the wide-$I$
bandpass if the $V-I$ colors of the reference stars have been measured
(i.e., for all L dwarf fields except 2M1658+70 and 2M2224-01), although the
calibrated correction must be extrapolated considerably for the mid- to
late-L dwarfs.  This correction allows for small, second-order DCR effects
resulting from the fact that the exposures are not centered {\it exactly}
on meridian transit.  Examination of the solutions carried out both with
and without the DCR correction reveals at most differences of a few tenths
of a mas in the derived relative parallaxes.  A DCR correction for
the $z^*$ filter has not yet been calibrated, although the effects are
expected to be smaller than the effects in the $I$ filter because atmospheric
dispersion is reduced.  Hence, particular care has been taken to examine
the parallax determinations for the four fields observed in $z^*$ by
carrying out solutions including a DCR term within the formal least-squares
algorithms and by examining the residuals to parallax results as a function
of the projected tangent of the zenith distance at the times of individual
observations (cf. Monet et al.\ 1992).  The effects of DCR are less than
0.3~mas in the parallax for these four objects.

\subsection{Present Results}

Table 1 presents the astrometric results for the 28 targets.  The first
column in Table 1 contains abbreviated names of the objects, also used
throughout the text.  The objects in Table 1
(as well as those in Tables 2 and 3 following) are presented in order of
increasing right ascension.  Full names, coordinates, and references for
all objects in Tables 1-3 are given in the Appendix.  The first column
also contains the spectral classification on the system established by
Kirkpatrick and co-workers (Kirkpatrick et al. 1991;
Kirkpatrick et al. 1995; K99; Burgasser et al. 2002).
The second column in Table 1 summarizes the data employed in each parallax
determination presented here: the total number of observations (CCD frames) and
the epoch range in years between the first and last observations.  The third
column gives the derived relative parallax and formal standard (mean)
error, while the fourth column gives the absolute parallax allowing for the
finite distance of the reference stars\footnote{
The formal least-squares reduction algorithms set both the mean parallax
and the mean proper motion of the reference stars employed in each field
identically equal to zero.  The corrections from relative to absolute parallax
employed here were derived from photometrically determined distances to the
individual reference stars using the same $VI$ photometry employed for the DCR
corrections.  As can be seen by comparing the entries in the third and fourth
columns, the mean distances of the reference frames are typically between
one and two kiloparsecs, and the uncertainty in this correction is only
rarely large enough to increase the uncertainty of the derived absolute
parallax.  For the six fields lacking reference star photometry to date
(the four $z^*$-filter fields plus 2M1658+70 and 2M2224$-$01),
a correction to absolute parallax of 0.8$\,\pm\,$0.2 mas has been adopted
based on similarly faint reference star frames located at comparable
galactic latitudes and with photometrically derived mean distances.
Note that all but one of these six targets (2M1658+70) is found to lie within
12 pc of the Sun, so that small uncertainties in the correction to absolute
parallax only compromise luminosity determinations derived from these parallaxes
at the 0.01 mag level.}.
The fifth and sixth columns give the total relative proper motion and position
angle of that motion (measured in the conventional sense of north through east)
obtained from the simultaneous least-squares solutions to the observations,
along with their formal uncertainties.  There is no simple and direct method
for correcting our derived relative proper motions to absolute (i.e., to a
quasi-inertial reference frame).
% Luyten employed the estimates of Paranago to
% convert his Palomar Survey measures to ``absolute" (Luyten 19xx).
% [NOTE: I'm still looking for the appropriate Luyten reference!]
For reference objects as faint as those employed for the majority of the
present targets, we expect such corrections to be less than a few mas per year.
Since most of the 28 objects in Table 1 have measured relative proper motions
in excess of 100 mas yr$^{-1}$ (the noteable exceptions being PC0025+04 and,
to a lesser extent, T513$-$46546 and 2M0326+29), the resulting uncertainties
for kinematic analyses are not large.  The seventh column contains the
tangential velocities (with respect to the Sun) calculated from the absolute
parallaxes and the relative proper motions given in the fourth and fifth
columns, respectively.

Most of the astrometric solutions presented in Table 1 are ``preliminary"
in the sense that these fields are still being observed on a regular basis.
The exceptions are CTI0126+28, LP412$-$31, T213$-$2005, T513$-$46546, and
GRH2208$-$20 which are no longer being observed and, therefore, the tabulated
determinations are considered ``completed."  Inspection of Table 1 reveals that
some fields have been observed for a long period of time -- over 9.1 years in
the case of PC0025+04 -- while others have been observed for only a short time
-- only 1.2 years in the case of SDSS1254$-$01.  Parallax determinations
from short time series of observations in particular must be regarded with
some suspicion.

\subsection{Reliability Checks}

We routinely employ two methods to judge the reliability, or robustness,
of our parallax determinations.  The first is based on experience gained
from the examination of over 230 TEK2K parallax solutions for over 220
different fields, including nearly a hundred with five years or more of
data coverage.  From this ensemble of solutions we know how the derived
parallaxes and their formal mean errors are expected to behave as a function
of time from the point that solutions are first possible (i.e., when the change
in Right Ascension parallax factor becomes non-monotonic) and then as data
frames are accummulated one by one.  These results have demonstrated that
(1) after about 2 years of observations individual solutions generally
remain very stable, and (2) the actual parallax value derived from that
point on agrees -- to within the formal error at any particular time  --
with the ``completed'' determination.  The formal error becomes smaller with
time as data are added to the solution, of course, and falls approximately
as N$^{-1/2}$, where N is the number of CCD frames in the solution.
Generally, the error in the parallax determination
at the time the star is removed as ``completed'' is some minimum value that
is approached quasi-asymptotically.  The actual value of this error is very
much field-dependent due to many factors including the quality of the reference
star field available and especially the coverage of the observations over the
parallactic ellipse.  In general, for a typical TEK2K field little improvement
in parallax precision is achieved with observations extending longer than
about five years.
 
As a second check we routinely examine and compare the two separate (but not
truly independent, since the same data frames are employed ) solutions for
parallax that are derived from image displacements in RA and DEC, respectively.
Maintaining these two solutions separately and only combining them into a
weighted mean relative parallax as a final step preserves useful diagnostic
information regarding contamination by (nearly) unresolved field stars,
DCR problems, or perturbations from unresolved physical companions.
While the range of parallax factor in DEC covered by the observations is
usually less (and frequently very considerably less) than realized in RA,
we should expect the two solutions to agree with one another to within the
combined errors of the two.  For all of the solutions presented in Table 1,
the individual RA and DEC parallax solutions agree satisfactorily -- that
is, usually to within their combined errors and no worse than 1.2 times that
amount.

Very preliminary parallax results were presented for 12 of the objects in
Table 1 at the ``Giant Planets to Cool Stars" conference held in June 1999
at Northern Arizona University (Dahn et al. 2000) and slightly updated values
were employed in Table 3 of K00.  Since that time some investigators have
expressed skepticsm about these values due to the lack of DCR corrections
for such extremely red targets (Mart\'{\i}n et al. 1999a) or due to the short
epoch coverage of the data ($\sim\,$1 year) for some fields.  As mentioned
above, restricting our observing close to times of meridan transit
effectively eliminates DCR as an issue.  Comparisons of the Table 1 values
with the earlier (1999) results demonstrate agreement that is at least
satisfactory and in most cases good to excellent, except for PC0025+04 and
DEN0205$-$11 which are discussed further in the following section.

Direct comparison of the present parallaxes with independent determinations
from other observatories are only available for three of the objects in
Table 1.  Tinney et al. (1995) presented parallax results based on CCD
observations from the Palomar 1.5~m reflector for these three objects.
For T832$-$10443, the agreement is excellent -- Pi(abs) =
36.4$\,\pm\,$3.0 from Tinney et al. versus Pi(abs) = 36.0$\,\pm\,$0.4 mas
from this paper.  For T513$-$46546 the agreement is less satisfactory --
Pi(abs) = 101.8$\,\pm\,$5.2 mas from Tinney et al. versus Pi(abs) =
93.2$\,\pm\,$0.5 from this paper.  However, the Tinney et al. determination
is based on observations spanning less than 1.6 years and the formal error
of their result is quite large.  The present result for T213$-$2005 --
Pi(abs) = 29.3$\,\pm\,$0.4 mas -- does not agree well the Tinney et al.
determination of Pi(abs) = 53.2$\,\pm\,$4.5 mas.  Tinney et al. note that an
anomalously large difference existed between their relative parallax solutions
in RA and DEC, and suggested that unresolved duplicity might explain this
discrepancy.  The Table 1 solution, which is based on observations covering
an epoch range of 7.0 years, shows no sign of a perturbation and the separate
RA and DEC solutions agree satisfactorily with one another
(29.0$\,\pm\,$0.4 mas versus 30.8$\,\pm\,$1.0 mas).

\subsection{Comments on Individual Targets}

The parallax determination for PC0025+04 has been particularly troublesome and
after 9 years of observations the solution remains quite fragile.  The formal
error for the relative parallax has still only come down to 1.6 mas.
Undoubtedly, this is due primarily to the faintness of this target.  Earlier
solutions yielded preliminary parallaxes for PC0025+04 from as high as 17 mas
down to about 13 mas, depending on what criteria were adopted for rejecting
lower-S/N data.  The solution presented in Table 1 employs the maximum data
possible; that is, only data frames very obviously bad have been rejected here.
The fact that the proper motion position angle of PC0025+04 ($\sim\,$95 deg)
is quite closely aligned in an east--west direction where the majority of
the parallactic shift occurs, and that the size of the annual proper motion
is comparable with the annual parallactic shift, undoubtedly contributes to
difficulties in simultaneously solving for these two quantities from low S/N
data, especially for observations extending over a short epoch range.
However, for observations covering 9 years, we would expect
the signals due to parallax and proper motion to be well separated by now.
The Table 1 parallax now places it above the loci of other points
in various absolute magnitude versus color diagrams (see Fig. 4 below),
suggesting the possibility that this is an unresolved binary system
(Burgasser et al.\ 2000a).
The residuals from our astrometric solution only allow us to rule out
perturbations with amplitudes larger than about 1 mas and for periods in
the 2 to 7 year range.  Shorter periods would not be resolved in our limited,
low S/N data.  Mart\'{\i}n et al. (1999a) discussed the status of PC0025+04
in detail and reported no radial velocity variability in observations on
nine nights covering 3.1 years.  However, they detected H$\alpha$ emission,
variable but consistently high (EW$\sim\,$100--400 \AA), and may have
detected LiI $\lambda\,$6708 \AA\ absorption (EW=1.0$\pm$0.3 \AA).
The absolute magnitude derived from our most recent astrometry seems
at least qualitatively consistent with
the Mart\'{\i}n et al. conclusion that PC0025+04 is probably a young
($<$1 Gyr) object.

The 1999 parallax solution for DEN0205$-$11 (59.4$\,\pm\,$2.6 mas) differs
quite significantly from the current (Table 1) value of 49.8$\,\pm\,$1.5 mas.
Examining the evolution of the solution to date reveals that, while the error
in the parallax has been decreasing with time, so too has the parallax itself.
DEN0205$-$11 is now known to be a close (Sep.$\,\sim\,$0.4--0.5 arcsec) binary
pair (Koerner et al. 1999), but since the components have equal brightnesses
at $JHK$ bandpasses (Leggett et al. 2001), a reasonable assumption is
that they also have equal brightnesses at the $I$-band employed for our
astrometric observations.  If this is the case, then the center of light
is very likely coincident with the center of mass of the system and
our parallax astrometry should not be affected.  Our observational coverage
of the parallactic ellipse is not particularly good for this field and this
may have been a problem for the 1999 solution whose observations included
only 1.2 years of epoch coverage.

The 1999 results for 2M0149+29 were based on an epoch range of only 0.4 years
and were presented to demonstrate that under favorable conditions
(i.e., intense observational coverage) meaningful results could be obtained
even when the change of the parallax factor in RA was still monotonic --
{\it if} that change were sufficiently non-linear with time.  The present
astrometry shows residuals in both the individual RA and DEC solutions
that hint at a possible perturbation.  However, it is premature to conclude
that this is an unresolved binary system.  Liebert et al. (1999) discuss the
variability in the spectra observed for 2M0149+29 and compare the H$\alpha$
activity with that observed for PC0025+04, concluding that
2M0149+29 is more like a single object which exhibits spectacular flaring.
For further discussion here we will assume that 2M0149+29 is single; but we
will continue to monitor the astrometry for evidence of a perturbation.

\section{PHOTOMETRY}

For very cool dwarfs, broad-band photometry at near-infrared wavelengths
is primarily sensitive to the strong molecular bands of H$_2$O,
CH$_4$, CO, and H$_2$.
Broadband photometry at ``optical'' wavelengths ($\lambda \leq 1 \mu$)
measures very different spectrophotometric features.
Despite the faintness of cool dwarfs, optical photometry is
potentially useful as diagostic of the stellar temperature,
surface gravity, metallicity, and the dust content of the photosphere
(e.g. Marley et al.\ 2002; Burrows et al.\ 2002).
In this section, we present both optical and near-IR photometry in order
to explore (in the following section) different color-magnitude diagrams,
seeking colors that are most sensitive to the physical properties and
evolutionary state of the objects.  New photometric data are presented
in Table 2.  Data from the literature are summarized in Table 3.

\subsection{New Optical Photometry}

We have chosen to measure $V$, $R_{\rm C}$, $I_{\rm C}$, and
$z^*$ where possible, and to transform data from the
literature to these passbands where practical.\footnote{
Here $V$ is on the Johnson system, $R_{\rm C}$ and $I_{\rm C}$
are on the Cousins system, and $z^*$ is on the SDSS system.
The $z^*$ (and other) SDSS magnitudes have zeropoints defined
to be on an AB magnitude system, while $VRI$ and the infrared $JHK$
data described below have zeropoints chosen to give Vega-like stars
colors of zero.  Combining these different magnitudes into colors like
$I_{\rm C}-z^*$ or $z^*-J$ is legitimate if it is done
consistently.}
These bands cover the red portion of the optical spectrum, and extensive
photometry of dM stars already is available in the $VRI$ filters.
The $I_{\rm C}$ filter is strongly affected by the KI resonance
doublet lines in late-L and T dwarfs, making $I-z^*$ and $I-J$ colors
even more sensitive to the steepening slope of the optical continuum
at late spectral types than they would otherwise be.
The $R_{\rm C}$ filter is also affected by the KI lines.
The $V$ filter is strongly affected by the NaI resonance lines.

Obtaining accurately-calibrated photometry for objects that are much
redder than any photometric standard stars requires some care.
Most of the measured flux in $VRIz^*$ filters for late-M, L,
and T dwarfs comes from the few hundred \AA\ at the red end of each
passband, whereas the measured flux for standard stars comes from
the entire passband.  Therefore, non-linear color terms are to be
expected when calibrating instrumental magnitudes.
Photometry taken with different photometric systems can be
transformed to other systems, but the corrections will be large when
the red cutoffs of the filters are very different.  For example,
SDSS $i^*$, Cousins $I_{\rm C}$, Thuan-Gunn $i$, and HST F814W are
similar filters, primarily measuring the flux redward of the KI lines,
but their red edges (at 8180, 8820, 8700, and 9600 \AA, respectively)
are sufficiently different that a late-L dwarf will be measured
differently by 2.4 mag between the extreme two of these four filters,
as we discuss below.

The new optical photometry presented in Table 2 includes all 28
objects from Table 1 plus 24 additional cool dwarfs with distances
already known or that will be useful in exploring the color--spectral
type diagrams.  The new photometry was obtained
primarily with the USNO 1.0~m telescope on 28 nights during 1998-2000.
Some $z^*$ observations of the faintest objects were made with the
USNO 1.55~m telescope.  Data (or lower limits) in $V$ and $R_{\rm C}$
for six objects (2M0850+10, 2M1328+21, 2M1553+21, 2M1632+19,
2M1726+15, and 2M2101+17) were obtained with the WIYN\footnote{
The WIYN Observatory is a joint facility of the University of
Wisconsin-Madison, Indiana University, Yale University, and the
National Optical Astronomy Observatories.}
3.5~m telescope on six nights in 1999.
All $VRI$ data were reduced using Landolt (1983, 1992) standards,
always including stars with a wide range of colors, typically
$V-I$ of 0.4-2.4 or greater range.
Nightly measurements were made for extinction and color terms.
For the faint program objects, sometimes only one filter was observed
on a given night, and the color terms were applied using observations
in another filter from a different night in a bootstrap procedure.
Transformations of the form
\begin{displaymath}
I = i - a_0 - a_1X - a_2(R-I) - a_3(R-I)^2
\end{displaymath}
(and similar for the other filters), where $X$ is the airmass,
were used to correct instrumental magnitudes to the standard system.  
The $V$ and $R$ glass filters and the $I$ interference filter used
at USNO are close to the standard passbands, but do not match exactly.
As mentioned above, the red cutoff of each passband is of primary
importance for the red program stars in this paper.  The red side of
the passbands for our filter/detector cameras are about 60~\AA\ too red,
200~\AA\ too blue, and about 250~\AA\ too red, respectively, compared
to the $V$, $R$, and $I$ passbands tabulated by \cite{bes90};
the color terms in the transformations account for these differences.
We find good systematic agreement with data
from \cite{bes90b} for M stars, although some stars have larger
differences than expected that could be due to stellar variability.
See \cite{mon92} for a comparison of USNO photometry with data from
other sources.  For L and T dwarfs, there is little other optical
photometry with which to compare.

The $z^*$ filter, CCD, and telescope used for most of the $z^*$
observations in this paper are the same used to set up the preliminary
SDSS standard-star network.  Observations presented here were reduced
using preliminary standard values for the standard stars taken from
Smith et al. (2002), applying nightly extinction values but no
color terms.  Therefore, these data should be on the standard-star
system with good accuracy despite a limited range of color for the
observed standard stars.  However, the SDSS system will ultimately
be defined by the survey camera on the SDSS 2.5~m telescope,
where the red side of the $z^*$ passband is defined by the sensitivity
of the six thick CCDs with $z^*$ filters in that camera.
The SDSS observations reported to date for L and T dwarfs in several
papers are on a preliminary system not yet tied to the system of
standard stars with color terms, as they eventually will be.
Measurements of the QE of the $z$-band CCDs in the SDSS survey camera
indicate that they do not have as much sensitivity at wavelengths
beyond 10500~\AA\ as had been expected for thick CCDs;
thus, fortuitously, they match the sensitivity of the CCD in the USNO
camera quite well.  Data in Tables 2 and 3 (see Sec. 3.3 below)
for five objects observed with both telescopes show this agreement
as well: the differences (USNO-SDSS) for SDSS0330$-$00, SDSS0413$-$01,
SDSS0539$-$00, SDSS1254$-$01, and SDSS1624$-$00 are $-0.05{\pm}0.05$,
$0.00{\pm}0.04$, $-0.03{\pm}0.02$, $+0.05{\pm}0.06$, and
$-0.13{\pm}0.06$, respectively.  Therefore, the $z^*$ data in Table 2
are taken to be on the preliminary SDSS $z^*$ system until a final
SDSS photometric system is established.

\subsection{New Infrared Photometry}

The $J-H$, $H-K$, and $K$ colors and magnitudes presented in Table 2
were obtained with the IRCAM imager at the 1.55~m Strand Astrometric
Reflector at USNO, Flagstaff Station. IRCAM employs a 256 x 256 HgCdTe
NICMOS III array pixelized at 0.54$^{\prime\prime}$~pix$^{-1}$.
The numbers of independent observations, on different nights, are given
in the last three columns. Each observation of a program object consisted
of three or four images, each with several minutes of integration,
taken at dithered positions of the telescope, even for the
brightest objects. On each night of observation between 10 and 22
standard stars were observed in order to determine nightly extinction
coefficients and color terms. The standards were selected from the
list of Elias et al. (1982) and thus place the resulting photometry
on the CIT system.  Most data were reduced to the standard system
in terms of the $J-H$ and $H-K$ colors and $K$ magnitude.  For a few
objects noted in Table 2 with noisy or missing $H$ or $K$ data,
the $J$ magnitudes were reduced and reported separately.  The IRCAM
$JHK$ filters are well-matched to the CIT standard system with mean
deviations from 1.0 in the transformation slopes from instrumental to
standard colors $<$1\% for $J-H$ and $H-K$. No $J-K$ color term was
ever found for the $K$ magnitude instrumental to standard value offset.

We add the caveat to the USNO photometry results that the standards used 
above only extend to red colors of approximately ($J-H$, $H-K$) =
(0.81, 0.36); roughly that of a main sequence mid- to late-M star.
For the redder late-M and L dwarfs measured here, colors have been
derived by extrapolation beyond the range of the previously derived
standard system. 

\subsection{Additional Photometry}

Many late-M, L, and T dwarfs also have $JHK$ photometry available in the
literature, particularly from 2MASS, that can be used to supplement the
data in Table 2.
The recently introduced MKO-NIR system (Tokunaga et al.\ 2002)
also has been used for observing L and T dwarfs (Leggett et al. 2002a).
A detailed comparison of these different systems is beyond the scope of
this paper.  In Figure 1, we make a comparison for late-M through L dwarfs,
comparing USNO data on the CIT system (Table 2) with 2MASS photometry
(left panels) and with MKO photometry (right panels).  These comparisons
do not include T dwarfs, which are expected to behave differently.
Both 2MASS and MKO filters have passbands different from the
CIT-system filters, so some systematic differences are to be expected
\footnote{Note that 2MASS data are reduced using CIT-system standard
stars, but without application of color terms, thus effectively making
the 2MASS photometric system unique.
}.
Carpenter (2001) finds only a very small difference between 2MASS and CIT
$J$ magnitudes, including several red standards (but not including L or
T dwarfs).  Figure 1 suggests that for late-L dwarfs the 2MASS $J$
magnitudes are systematically fainter than the CIT $J$ magnitudes;
the five reddest objects have a mean difference of 0.14$\pm$0.06,
formally significant, but obviously not well-determined.
Because many L dwarfs have been discovered with
2MASS data using a selection for objects with $J-K$ colors redder than
some limit, a bias may exist in the $J$ and $K$ 2MASS magnitudes for
those faint L dwarfs discovered by 2MASS (K00).  Therefore, some of
the difference in $J$ magnitudes seen in Figure 1 may be a result of
this selection procedure.  Because this difference is small and of an
uncertain amount, no corrections to 2MASS data have been applied in
the following section.
However, the MKO system is significantly different from both the 2MASS
and CIT systems, resulting in color corrections for M, L, and T dwarfs
(Leggett et al.\ 2002a).  We therefore avoided its use in our analysis.

Table 3 gives the additional data used, where a weighted
mean has been adopted for objects with data from more than one source,
including objects listed in both Tables 2 and 3.
For near-infrared data, we made use of 2MASS photometry from K99, K00,
Gizis et al.\ (2000), Burgasser (2001), and Wilson et al.\ (2001).
The comparison of 2MASS and CIT photometry was described above.
% For 2MASS 0825+21, 2MASS 1237+6526, Gliese 584C we used weighted means
% between the 2MASS magnitudes and our USNO measurements to improve precision.
We also included UKIRT photometry (not on the MKO-NIR system;
see Leggett et al.\ 2002a) in our adopted values for the two T dwarfs
SDSS 1346$-$0031 ($K$ data only, from Tsvetanov et al.\ 2000;
$J$ and $H$ adopted values are on the CIT system listed in Table 2)
and Gl~229B ($JHK$ data, from Leggett et al.\ 1999);
UKIRT-system values are expected to be somewhat different from the
CIT and 2MASS values used elsewhere in this paper, but are the only
data presently available for these two stars.
% For GD 165B we combined our USNO JHK photometry with that of
% Becklin \& Zuckerman (1988) and Tinney et al.\ (1993), all obtained
% on the CIT system; we also incorporated the
% $I_{\rm C}$ measurement of Tinney et al.\ (1993).  To supplement our $z^*$
% photometry, we used published SDSS $z^*$ values for SDSS 0539$-$0059
% (Leggett et al.\ 2000), SDSS 1346$-$0031 (Tsvetanov et al.\ 2000),
% and SDSS 1624+0029 (Strauss et al.\ 1999).
$I_{\rm C}$ photometry for 2M1217$-$03, 2M1237+65,
SDSS1346$-$00, SDSS1624+00, 2M1726+15, 2M2101+17,
and Gl~570D was synthesized from spectral data from K00
and Burgasser et al.\ (2000b; 2001), and $z^*$ photometry
for Gl~570D was synthesized from combined spectral data from
Geballe et al.\ (2001) and Burgasser (2001).

Spectrophotometric colors for Gl~229B at $I_{\rm C}$ and $z^*$ could
not be directly obtained due to the short wavelength cutoff of the
Oppenheimer et al.\ (1998) spectrum (flux calibrated by Leggett
et al.\ 1999).  Therefore, we obtained this photometry by applying
color corrections to the Gunn $i$ and $z$ measurements from Nakajima
et al.\ (1995) and HST F814W and F1042M measurements from Golimowski
et al.\ (1998).  These corrections were synthesized from optical and
near-infrared spectra of mid-type T dwarfs from Burgasser (2001) and
Geballe et al.\ (2002).  We were careful to include the quantum efficiency
(QE) of the CCD detectors used for the photometric measurements,
particularly in the $z$ band where the CCD QE decay at long wavelengths
cuts off the otherwise long-pass Gunn and Sloan filters.  We derive
$I_{\rm C}$ = 20.5$\pm$0.3\footnote{Uncertainties are derived from the
original measurements and scatter in the color correction.}
and 19.92$\pm$0.14 from the Gunn $i$ and F814W magnitudes, respectively,
which are marginally consistent within the uncertainties.  We adopt
the HST-derived magnitude value for the remainder of the paper.
For $z$, we derive $z^*$ = 20.0$\pm$0.3 and 17.69$\pm$0.07 from the
Gunn $z$ and F1042M magnitudes, which are clearly inconsistent.
The mean Gunn $z$-F1042M color measured from the spectra of SDSS
1624+0029 and SDSS 1346-0031 (both classified T6; Gliese 229B is
classified T6.5 by Burgasser et al.\ (2002)) is 1.54$\pm$0.13,
as compared to 3.8$\pm$0.3 from the photometry.  Because the HST
F1042M-F814W color for Gl 229B is consistent with the spectral data,
we conclude that the Gunn $z$ measurement from Nakajima et al.\ (1995)
is descrepant by over 2 mags, possibly due to its significant
sensitivity to the properties of the CCD detector.
We therefore adopt the HST-derived value for $z^*$.
This exercise again emphasizes the need for careful photometric
calibration of L and T dwarfs at these wavelengths, due to
their significant spectral slopes at $\lambda\lambda$8000-11000 {\AA}.

\section{DISCUSSION}

\subsection{Absolute Magnitude--Spectral Type Relations}

The parallax values in Tables 1 and 3 and the photometry
in Tables 2 and 3 give the absolute magnitudes in each passband.
Five objects are known to be resolved binaries.\footnote{
One additional object, 2M0345+25, was suggested to be a double-lined
spectroscopic binary (Reid et al.\ 1999), but reexamination of the data
indicates that that the object is more likely single with broad lines.
}
For these binaries, photometry in the tables is given for the combined
light, so the magnitudes for the individual components have been
calculated as follows.  For two objects (DEN0205$-$11, and DEN1228$-$15),
the components are observed to have equal or nearly equal magnitudes at $JHK$
(Koerner et al.\ 1999; Mart\'{\i}n et al.\ 1999b; Leggett et al.\ 2001).
They are assumed to be identical objects with each component being
0.75 mag fainter than the combined light at all bandpasses.
For three other dwarfs (2M0746+20, 2M0850+10, and 2M1146+22),
the magnitude difference between the components has been measured
with the F814W filter on HST and the magnitude difference in $J$
has been estimated by \cite{rei01b}. 
Using the observed color-color plots for L dwarfs described below, we
have made estimates of the magnitude difference in other passbands,
including error estimates allowing for the scatter seen in $JHK$ colors
for L dwarfs and the unknown (late-L or T) spectral type of 2M0850+10B.
The estimated magnitude differences are listed for these three
binaries in Table 4.  From these values and the photometry for the
combined light given in Tables 2 and 3, the magnitude of each component
in each filter can be calculated, and is used for plotting in the figures.

Figure 2 presents $M_I$, $M_J$, and $M_K$ as a function of spectral type.
Spectral types are on the system established by Kirkpatrick and co-workers,
as described in Sec. 2.2.  They are accurate within that system to
typically one-half spectral subtype or better.
All three diagrams show clear monotonic relations from mid-M through
late-L with a nearly linear slope.
There is some scatter in these relations with a range of
$\pm$0.5 mag; this scatter is significantly greater than the
measurement uncertainty and, hence, is likely due to both cosmic scatter
(age, metallicity, etc.)  and possible binarity.
There is a clear break in this relation as we cross into the
T dwarf regime, however.  Both SDSS1254$-$01 (T2) and 2M0559$-$14
(T5) have $M_I$ and $M_K$ values that are similar to the latest L dwarfs,
and 2M0559$-$14 is more than a magnitude brighter at $J$ than the
L8 dwarfs.  Even if this object is an equal-magnitude binary (despite
it being unresolved in HST images -- Burgasser 2001),
then both $M_I$ and $M_K$ show a moderate dimming from L8 to T5,
while $M_J$ shows a moderate brightening from L8 to T5, followed by all
magnitudes rapidly dimming for late-T dwarfs to Gl~570D at T8.
This pattern could be consistent with substantial flux redistribution
toward $J$ as these objects cool, causing $M_J$ to brighten temporarily
even as the bolometric luminosity is decreasing (see Sec. 5).
Alternatively, the unexpected bright magnitudes of 2M0559$-$14 may be a
symptom of more complicated behavior in early-T dwarfs, and a real dispersion
in luminosities as large or larger than that seen in L dwarfs.
Distances for more T dwarfs are urgently needed to address this issue.

\subsection{Spectral Type--Color Relations}

In Figure 3, a selection of optical colors are plotted as a function of
spectral type, and Figure 4 shows $I-J$ and $J-K$ against spectral type.
In general, the optical colors grow redder with increasing spectral type.
The $R-I$ and $I-z^*$ colors are nearly constant from M8 through L4
spectral types, as has been previously noted (Fan et al. 2000; 
Steele \& Howells 2000; Schneider et al. 2002) while $V-I$ and $z^*-J$
are increasing.  With a few exceptions, $I-z^*$ shows very little scatter
at a given spectral type.  Notable
outliers are PC0025+04 (M9.5) and 2M1841+31 (L4), both with peculiar
spectra, and GD165B (L4) which may have its $I$ magnitude corrupted by
its brighter primary.
Probably $R-I$, $I-z^*$, and spectral type are all heavily influenced by
the relative prominance of molecular bands (TiO and VO) vs. atomic (KI)
absorption in late-M through mid-L dwarfs.
$I-J$ appears to be good diagnostic of spectral type, increasing
monotonically from mid-M to late-T, but with more scatter than $I-z^*$.
There is a change in slope between the late-M dwarfs and L dwarfs,
with $I-J$ colors increasing less rapidly in the latter objects.
In the latest L dwarfs and T dwarfs, $I-J$ color is heavily
influenced by KI absorption, and the increased reddening is
consistent with the observed (Burgasser 2001) and predicted
(Burrows et al.\ 2002) strengthening of the
$\lambda\lambda\,$8000--11000 {\AA} slope for these spectral types.
As discussed in Burrows et al.\ (2002), the strength of KI absorption
is also heavily dependent on gravity and metallicity, a fact that may
explain the significant scatter seen in some L and T dwarfs.

$J-K$ color trends in M, L, and T dwarfs have been examined by a number
of authors, and we see similar behavior in the bottom, right-hand panel
of Figure 4:
reddening from M to L, likely caused by dust, followed by a significant
change to blue near-IR colors in the T dwarfs where CH$_4$, H$_2$O, and
CIA H$_2$ opacity dominate.  Note the significant scatter in colors between
spectra types L2 and L8.  Leggett et al.\ (2002a) attribute this behavior
to variations in the amount of dust present in the photospheres of these
objects.  We concur that the scatter (which is larger than our photometric
uncertainties) must be intrinsic to the objects themselves.  This scatter
is also apparent in $z^*-J$ colors, but not in $I-z^*$.  Note that
one object, the L6.5 2M2244+20, is nearly 0.5 mag redder than
all other L dwarfs and appears to be an especially peculiar object.
There may also be scatter in the colors of mid- and late-T dwarfs,
indicated in the $zJHK$ data in Leggett et al.\ (2002a) and
Burgasser et al.\ (2002),
possibly due to variations in H$_2$ opacity (Burgasser et al.\ 2002).
The color uncertainties in this paper are generally too large and the
sample too small for us to say more about these very cool brown dwarfs.

\subsection{Color--Absolute Magnitude Relations}

Figure 4 also shows the absolute magnitude $M_J$ plotted against $I-J$ and
$J-K$.  The $M_J$ versus $I-J$ relation shows a tight sequence for late-M and
L dwarfs, aside from a few notable outliers that are marked.  The M and L
dwarfs lying to the right of the sequence may be unresolved binaries or may
be young.  The object to the left of the sequence (GD165B) may have an error
in the $I-J$ color caused by contamination from its bright companion.
A simple formula reproduces this sequence:  for 37 dwarfs with
spectral type M6.5 through L8 (excluding the four outliers PC0025+04,
2M1328+21, GD165B, and LHS102B),
\begin{displaymath}
M_J = 21.81 - 8.692(I-J) + 1.697(I-J)^2.
\end{displaymath}
This formula is valid for $2.8 < I-J < 4.2$.
The rms dispersion is only 0.23 mag from this relation, hardly larger
than the observational errors.
The dispersion here is slightly smaller than was observed for $M_J$
versus spectral type in Figure 2;  there, a similar fit for 45 objects
with spectral type M6.5 through L8 (excluding the two greatest outliers
LP944$-$20 and 2M0850+10) is
\begin{displaymath}
M_J = 8.38 + 0.341*ST
\end{displaymath}
(where $ST$ = 7 for spectral type M7 up to 18 for spectral type L8),
with a dispersion of 0.25 mag.  For the early-T dwarfs SDSS1254-01 and
2M0559-14, $M_J$ appears to be as bright or brighter than for late-L
dwarfs, similar to the behavior seen in Fig. 3.  A redistribution of flux
from the mid-IR into the $J$ band may help to push the $I-J$ colors of
T dwarfs 1--2 magnitudes redder than the L dwarfs (see Sec. 5).

The right panels of Figure 4 show that the scatter in the $J-K$
colors of mid- and late-L stars when plotted against $M_J$ persists,
just as the scatter appeared when plotted against spectral type.
The scatter in absolute magnitude was
noticed before (K00), but was somewhat tentative because of the
few L dwarfs with measured distances.  Figure 4 shows that there is
a trend {\it in the mean} $J-K$ colors of L dwarfs, but the
correlation between $J-K$ and $M_J$ for individual stars is very poor.
Because $M_J$ is closely coupled with effective temperature for M and L
dwarfs (Sec. 4.7 below), Figure 4 shows that some factor in addition to
temperature is dominating the $J-K$ colors for L2-L8 dwarfs.  This puzzle
is discussed further in Section 5 below.

\subsection{Color--Color Relations}

Figure 5 plots a selection of color-color diagrams for all of the
objects with measured optical and near-infrared photometry.
They show the utility of the $z^*$ filter both for classification
purposes and for deriving physical parameters.
Spectral types and subtypes can be accurately estimated using the
combination of $Iz^*JK$ photometry.  Even $z^*$, $J$, and $K$ alone
provide an approximate classification, although the scatter in the colors
of mid- to late-L dwarfs makes their classification less accurate
without additional colors.  (Using $z^*JK$ alone is useful for classifying
faint dwarfs with SDSS $z^*$ data plus 2MASS or other $JK$, but without
$r^*$ or $i^*$ data.  The same technique should be useful for classifying
objects with $IJK$ from DENIS.) The L6.5 dwarf with very red $J-K$ color
(2M2244+20) clearly merits further study.  A parallax determination
for this unusual object will be particularly interesting.

The first panel ($J-K$ vs. $I-z^*$) is similar to one studied by Marley
et al. (2002), who showed that $i^*-z^*$ and $J-K$ colors are sensitive
to the presence of condensates in the atmospheres of cool dwarfs.
Cloud-bearing models with sedimentation provide a qualitative understanding
of this diagram.  The significant scatter and lack of
correlation among L dwarfs suggests that different values of the
rainout efficiency parameter (see Ackerman \& Marley 2001 for details
on this model) might be needed to match the data.  For T dwarfs,
the models predict that $J-K$ will decrease while $I-z^*$ increases,
until a maximum is reached in $I-z^*$ for objects that have cooled to
between $-0.5 > J-K > -1.5$ (depending on rainout efficiency parameter).
Our data are roughly consistent with these predictions, although the
redder $I-z^*$ color of Gl~570D might suggest that we have not yet reached
the turn around point.  However, the $I-z^*$ color for Gl~570D is based
on two synthesized magnitudes, and it is also possible that the derived
color is more uncertain than is estimated here.

\subsection{Kinematics}

Tangential velocities of stars comprise two of the three components of
their space velocities and, as such, should provide some statistical
indication of a population's age.  Radial velocity measurements have
appeared in the literature for a few stars in this paper,
and for them space velocities can be determined.
However, for most of the sample, only the tangential velocities
are available.  A discussion of $V_{\rm tan}$ for a few early- to mid-L
dwarfs showed the data were consistent with them having ages
${\sim}1$~Gyr or older (Gizis et al. 2000).
A comparison can be made with the tangential velocities in Table 1 with
those of the volume-complete sample of M dwarfs (primarily M0 to M5)
in the solar neighborhood from \cite{rei95} and \cite{haw96}.
The dM stars have $<V_{\rm tan}> \simeq 44$~km~s$^{-1}$ for all absolute
magnitude bins from $M_V = 8.5$ to 14.5 (spectral types M0 to M5),
and 27\% of the stars have $V_{\rm tan} \ge 60$~km~s$^{-1}$.
The subset of dMe stars have somewhat smaller motions due to their younger
mean age, with $<V_{\rm tan}> = 31$~km~s$^{-1}$.
The objects in Table 1 have $<V_{\rm tan}> = 34 \pm 5$~km~s$^{-1}$
and four objects (14\%) have $V_{\rm tan} \ge 60$~km~s$^{-1}$.
These results are intermediate between the values for all dM stars
and the dMe stars.  Comparison with a relation between age and velocity
dispersion for disk stars (e.g. Wielen 1977) indicates that the sample
has a mean age of 2-4 Gyr.
One star (GRH2208$-$20) has a moderately-high velocity, suggesting
(but not requiring) thick-disk or even halo membership;  if it is
excluded, the remaining sample has $<V_{\rm tan}> = 31 \pm 4$~km~s$^{-1}$
There is no obvious difference between the M, L, and T dwarfs in Table 1.

These results show that the objects in this paper have slightly smaller
space motions compared to the earlier-type M dwarfs, although the difference
is of marginal significance.  A very similar result has been found
by \cite{rei02} for a large sample of late-M dwarfs (M7 to M9.5).
A difference like this could be caused by a Malmquist bias ---
the preferential discovery of more luminous L and T dwarfs
that would be systematically younger and have smaller velocities.
However, the small dispersion in absolute magnitudes seen in Figure 4,
$M_J$ vs. $I-J$, would indicate that not much dispersion exists among
most late-M and L dwarfs to affect the kinematics of the sample.
\cite{rei02} drew the same conclusion for their late-M dwarfs.
Alternatively, smaller space motions for L and T dwarfs compared to M
dwarfs can be understood as a result of a combination of two factors.
First, late-L and T dwarfs encompass a wider range of masses at a
given effective temperature than do M dwarfs, as can be seen from
evolutionary models (e.g. Chabrier et al. 2000) over a range of ages.
Second, a rising IMF boosts the number of lower-mass and younger objects.
This second factor becomes more important if the IMF is steep, but is
significant even with a slowly rising IMF (e.g. $\alpha \sim 1$).
A simple population model with these factors predicts a drop in space
velocities in agreement with the observed values, and predicts a further
drop for T dwarfs as long as the IMF continues to rise for lower masses.
The reality of this difference in motions between the M0-M5 dwarfs and cooler
dwarfs, and its cause, can be studied further when larger samples of
L and T dwarfs are available.

\subsection{Stellar Models and Ages}

The color-absolute magnitude diagram for M and L dwarfs shown in
Figure 4 is repeated in Figure 6, with evolutionary models taken from
\cite{cha00} using DUSTY model atmospheres (Allard et al. 2001)
added for comparison with the data.  T dwarfs are omitted
because the DUSTY models are known to be inappropriate for T dwarfs,
where the photospheres are dominated by molecular opacity, not dust.
The models with ages consistent with the kinematics of the sample
(ages of several Gyr) match the data qualitatively, but significant
differences are apparent:  for late-M and early-L stars, the models
predict optical and near-IR colors bluer than are observed;
for late-L dwarfs, the models predict near-IR colors much redder
and optical colors slightly redder than are observed.
These differences have been noted previously (e.g. Leggett et al. 2000a;
Leggett et al. 2001), as have other symptoms of models that are not
totally correct for these very challenging cool atmospheres (e.g. Reid
\& Cruz 2002).  The differences have been attributed, at least in part,
to incorrect opacity tables for H$_2$O used in the model atmospheres for
late-M dwarfs, and the unrealistic assumption that all dust remains in
the atmosphere of late-L dwarfs.  Models in which dust is distributed
in stratified clouds and is below the photosphere in late-L and T dwarfs
(e.g. Marley et al.\ 2002; Allard et al. in preparation) should provide
a much better match to the data.  Meanwhile, the empirical fit determined
in Sec. 4.3 and plotted in Figure 6 describes the sequence of disk L dwarfs
much more accurately than the dusty models.

\subsection{Temperatures of Late-M and L Dwarfs}

Effective temperatures can be calculated from the absolute magnitudes
from this paper if the bolometric corrections and stellar radii can be
determined.  This method of determining temperatures is potentially
very accurate for the low-mass objects in this paper, because only the
radii are model-dependent, and the radii depend primarily on the
interior models which are relatively well known.  Deficiencies in the
model atmospheres, such as discussed in the previous section, have
little effect on the derived temperatures using this approach.
This method was first applied to an L dwarf to estimate the temperature
of 2M1523+30 (Gl~584C) by K00, and has been used by others
(e.g. Leggett et al.\ 2001; Leggett et al.\ 2002a;
Burgasser et al.\ 2002).  Here we repeat the procedure for the late-M
and L dwarfs in this paper with known distances.

Bolometric corrections for $K$-band (BC$_K$) have been calculated by
\cite{leg01,leg02a} and references therein for M and L dwarfs.
Values for BC$_J$ have been calculated by \cite{rei01a},
and they agree very well for M through mid-L dwarfs, but are up to
0.4 mag different for L7.5-L8 dwarfs.  The difference comes from
Leggett's use of $L$-band photometry and assuming a Rayleigh-Jeans
flux distribution at longer wavelengths, thus deriving a larger
luminosity and smaller values for BC$_K$ and BC$_J$ than the values
derived by Reid et al.
The different results point out our uncertainty in the flux emitted
longward of 3 microns by these very cool objects.  Nevertheless,
because the temperature depends on the fourth root of the luminosity,
temperatures can still be derived despite this uncertainty.
An interpolation formula for BC$_K$ vs. $I-K$ was derived by
Leggett et al. for M and L dwarfs, and is adopted here.
(A formula for BC$_K$ vs. $J-K$  was also derived,
but is not as well established for late-L dwarfs,
with their wide range of $J-K$ colors, so is not used here.)  
This formula predicts values for BC that are 0.1 mag larger than
calculated by \cite{leg02a} and 0.3 mag smaller than calculated by
\cite{rei01a} for L7.5-L8 dwarfs.

Stellar radii for models from the Lyon group (Chabrier et al.\ 2000)
and from the Arizona group (Burrows et al.\ 1997) have a small systematic
difference throughout the regime of intermediate-age and old L dwarfs:
the models from the Lyon group have radii consistently larger by 6\%
at a given age and luminosity than those from the Arizona group.
Here we adopt an interpolation formula
\begin{displaymath}
R/R_{\odot} = 0.088 + 0.00070(16.2-M_{\rm bol})^{2.9}
\end{displaymath}
that predicts radii half way between the 1~Gyr and 5~Gyr models from
the two groups over the range 12-16.5 in $M_{\rm bol}$.
Therefore, the temperatures derived here would be higher if one were
to adopt radii from the Arizona models, and lower if one were to adopt
radii from the Lyon models, but the differences would be only 40~K
for M dwarfs and 20~K for late-L dwarfs.
We estimate an error in BC$_K$ of $\pm$0.05 mag at $M_{\rm bol}$~=~12
increasing to $\pm$0.10 mag at $M_{\rm bol}$~=~16, and an error
in $R/R_{\odot}$ of $\pm$3-6\% at $M_{\rm bol}$~=~12-16.

Table 5 contains the derived effective temperatures for the M and L
dwarfs in Tables 1 and 3 with the requisite data (parallax and
$I_{\rm C}JK$ photometry).  Selected relations based on these
temperatures are presented in Figure 7.
Use of bolometric corrections from \cite{rei01a}
would result in temperatures cooler by 100~K for the coolest three stars.
For other objects, the estimated BC$_K$ could
have a systematic error, but plausible systematic errors are included
in the error bars and can be seen to be quite small.
The temperatures derived here are hotter for late-M and early-L dwarfs
and cooler for late-L dwarfs than those derived in some other studies
(Basri et al. 2000; Leggett et al.\ 2001; Schweitzer et al.\ 2001;
Schweitzer et al.\ 2002; Chabrier 2002)
using fitting of model atmospheres to spectroscopic data.
These differences are probably further manifestations of the imperfect
models used in those studies, discussed in the previous section.
The mean temperature derived here for the three latest L dwarfs
(L7.5-L8) is 1360~K.  The relation between $T_{\rm eff}$ and spectral
type for L dwarfs shown in the bottom panel of Figure 7 is very similar
to the relation derived by Stephens et al.\ (2001) using $K-L$ colors
and the models of Marley et al.\ (2002).

The scatter in the bottom panel of Figure 7
is larger than might be expected, given the tight relations between
$M_J$, $I-J$, and spectral type seen in Figure 4.
(The range of $M_J$ at a given spectral type seen in Figure 2 is
another symptom of the scatter in Figure 7.)
Some objects (e.g., T513$-$46546, LP944$-$20, 2M0850+10)
are cooler than the mean for their spectral type, whereas others
(Kelu-1, 2M1112+35, DEN0205$-$11) are warmer.
The range in temperature of 200-300~K at a given spectral type is
somewhat larger than might be expected from the formal errors.
Unresolved binaries (in addition to the
binaries already noted in the tables and accounted for in Figure 7)
will contribute to this scatter, because they will have measured
luminosities too large (by up to a factor of two), adopted radii
too large (by up to 10\% for late-M dwarfs, less for L dwarfs),
and derived temperatures too hot by up to 240~K.
However, the small scatter in Figure 4 makes it unlikely that many of
the objects are unresolved binaries, enough to account for the range
of temperatures seen in the bottom panel of Figure 7.  Furthermore,
a few objects may be significantly younger (ages ${\leq}10^8$~yr),
have radii larger than assumed, and their derived temperatures will
be too high.  Much of the observed scatter could be accounted for by
revisions in the spectral types assigned to a few stars.  This
possibility would require changing the type of T513$-$46546 to a later
type of L0 (instead of M8.5), LP944$-$20 to L2 (not M9), and
2M0850+10 to L7-L8 (not L6); and if they are not binary (triple or
greater, in the case of DEN0205$-$11!) or young,
revising Kelu-1 to an earlier type of L0-L1 (not L2), 2M1112+35 L3
(not L4.5), and DEN0205$-$11 L6 (not L7).  Such revisions in spectral
type would be completely inconsistent with the morphology of the
optical spectra for at least some of these well-studied objects.
Using spectral types from \cite{mar99c} or \cite{geb02} would not
reduce the observed scatter in this diagram.
Alternatively, spectral types may be affected to a relatively minor
extent by some other physical parameter(s) in addition to temperature,
thus leading to an imperfect correlation with temperature that is
seen in Figure 7.

\section{CONCLUSIONS}

New or improved parallax measurements have been presented for
28 late-type dwarfs, including 17 L and 3 T dwarfs.
Optical ($VRIz^*$) and near-IR ($JHK$) photometry is also presented
for these and other cool dwarfs.  Various colors are used to
explore spectral type vs. color vs. magnitude diagrams.
Enough L dwarfs are included to begin to define their absolute
magnitudes reasonably well, but a larger sample of T dwarfs is needed.
The $I-J$ color is a good predictor of absolute magnitude for
late-M and L dwarfs, spectral type is almost as good, but $JHK$
colors are notably inferior for predicting absolute magnitude.

While only one L/T transition object has a measured parallax,
it and one of the mid-T dwarfs
both show $M_J$ brighter than for the latest L dwarfs.
This brightening of $M_J$ is accompanied by an increase in $I-J$
and a decrease in $J-K$, all indicative of an increase in the
fraction of total flux emitted in the $J$ band for early-T dwarfs
compared to late-L dwarfs, but with little difference in luminosity.
Apparently the early-T dwarfs have less dust above the photosphere,
or perhaps holes have developed through high-level dust clouds,
leading to increased molecular absorption at mid-IR wavelengths,
a redistribution of flux into the $J$ band, and the observed changes
in colors.  This brightening for early-T dwarfs would then be a
temporary phenomenon following the dusty L-dwarf phase and
prior to cooling further to late-T spectral type.  This picture
is consistent with the precipitation condensates described by
\cite{ack01}, for example.  If confirmed with a larger sample,
this behavior will affect the space density derived for these objects.

The combination of $z^*JK$ colors alone can be used to accurately
classify late-M, early-L, and T dwarfs, and predict their absolute
magnitudes, but is less effective at untangling the scatter among
mid- and late-L dwarfs.  The tangential velocity is derived for each
object, and the mean velocity is slightly less than that for dM stars
in the solar neighborhood, consistent with a sample with a mean age
of several Gyr.  Using colors to estimate bolometric
corrections, and models to estimate stellar radii,
effective temperatures are derived for the late-M and L dwarfs
with known distances.  The latest L dwarfs are found to have
$T_{\rm eff} \sim 1360$~K, with a range of 1250-1400~K allowed
by uncertainties in the bolometric corrections.

The $J-K$ colors of L dwarfs are thought to be affected by the
distribution of dust in the atmosphere.  Objects with dust distributed
high in the atmosphere should have redder $J-K$ colors, more in
agreement with the DUSTY models described in Figure 6.
In contrast, objects with atmospheres partially clear of dust
(perhaps because sedimentation or precipitation dominates over
turbulent mixing in the upper atmosphere) should have stronger
molecular absorption and $J-K$ colors that are not as red.
This behavior is shown in the models of \cite{mar02} as their
rainout parameter is changed.  This picture seems plausible,
but it is not yet clear what underlying physical parameters cause
the different behavior among objects that otherwise appear similar.
A factor that could be important is rotation velocity.
The nine L dwarfs that have measurements of $v {\rm sin} i$
(Basri et al.\ 2000; Reid et al.\ 2002)
all show a significant rotation, but there is not an obvious
correlation between the measured $v {\rm sin} i$ and the $J-K$ color.
For example, Kelu-1 and DEN1058-15 both have fast rotation, but Kelu-1
is redder than the mean for its absolute magnitude, while DEN1058-15
is bluer.  Kelu-1 may be atypical for a field L dwarf, and its high
rotation velocity and/or its red $J-K$ color may be related to its
variability (Clarke et al.\ 2002).  Measurements of  $v {\rm sin} i$
for additional L dwarfs may help reveal whether a correlation exists.
Of course, the distribution of dust could be quite sensitive to small
effects, and more than one factor could be contributing to the scatter
seen in L dwarfs so as to mask any simple correlation.
Parameters such as metallicity and magnetic field strength come to mind
as plausible factors to affect the dust content of these objects.
We consider the puzzling behavior of the near-IR flux of L dwarfs
to be an important problem for understanding these objects.

Parallax measurements for additional cool dwarfs are desirable,
especially for T dwarfs, for further understanding of their properties.
The faintness at optical wavelengths of late-L and T dwarfs
precludes observing substantial numbers of them with moderate-size
telescopes and CCD cameras.  The astrometric satellites being planned,
even those like GAIA and SIM that are designed to observe faint targets,
are unlikely to help for the same reason.  We will continue the present
CCD parallax work and may include some of the brightest objects as they
are discovered, but observing many objects this way is very time
consuming.  Further progress will require larger telescopes
and/or observing at near-IR wavelengths where the targets are brighter.
The USNO IR group has commissioned a camera employing an ALADDIN InSb
array on the Strand 1.55~m telescope, and is now one year into an
exploratory program of measuring parallaxes with $J$ and $H$ filters.
This program currently includes 13 L dwarfs with spectral type L6
or later and 16 T dwarfs, in addition to the three T dwarfs in Table 1.
Results from this program should be a first step in providing the
substantial numbers of late-L and T dwarfs with parallax measurements
that are needed.

\acknowledgments

We thank the referee, Sandy Leggett, for a careful reading of the manuscript
and for several useful suggestions for clarifying the presentation.
We thank our colleagues in SDSS for alerting us to their discovery
of the two T dwarfs prior to publication.
We thank the WIYN Queue staff for taking photometric data of some of
these objects with the WIYN telescope.
This publication makes use of data products from the
Two Micron All Sky Survey, which is a joint project of the
University of Massachusetts and the Infrared Processing and Analysis
Center/California Institute of Technology, funded by the
National Aeronautics and Space Administration and the
National Science Foundation.
This research has made use of the Simbad database,
operated at CDS, Strasbourg, France.

% \appendix
\section{APPENDIX}
Column 1 of Table A1 gives the full names of the stars listed in Tables
1, 2, and 3 in the main body of this paper.  Additional data for each
object is given as extracted from the literature:
the second and third columns give the approximate J2000.0 coordinates,
the fourth column gives a reference for the coordinates, and the fifth
column gives a reference where an identification chart can be found.

\clearpage

\figcaption[]{Differences between $JHK$ magnitudes of late-M and
L dwarfs observed in the CIT system (data in Table 2 of this paper)
and in the 2MASS system (left panels, data from various sources)
and in the MKO system (right panels, data from Table 3 of
Leggett et al. 2001b).  T dwarfs will behave differently, and are not
included.
\label{fig1}}

\figcaption[]{Absolute magnitudes of cool dwarfs with known distances
plotted against spectral type.  In this and following plots,
M dwarfs are plotted with squares,
L dwarfs are plotted with circles, and
T dwarfs are plotted with triangles.
\label{fig2}}

\figcaption[]{Selected colors of cool dwarfs plotted against spectral type.
Squares, circles, and triangles are M, L, and T dwarfs, respectively.
\label{fig3}}

\figcaption[]{Absolute J magnitudes and spectral types of cool
dwarfs plotted against $I-J$ and $J-K$ colors.
Squares, circles, and triangles are M, L, and T dwarfs, respectively.
\label{fig4}}

\figcaption[]{Selected color---color diagrams of cool dwarfs showing the
utility of the $z^*$ filter.
Squares, circles, and triangles are M, L, and T dwarfs, respectively.
\label{fig5}}

\figcaption[]{M and L dwarfs compared to the AMES DUSTY models
(Chabrier et al. 2000) for three different ages.  The solid line
shows a fit to the observed M6.5-L8 dwarfs (Sec. 4.3).
Squares and circles are M and L dwarfs, respectively.
\label{fig6}}

\figcaption[]{Effective temperatures of M and L dwarfs calculated
from $M_J$ (this paper), BC$_K$ (Leggett et al. 2001a), and
$R/R_{\odot}$ (models from Burrows et al. 1997 and Chabrier et al. 2000).
Squares and circles are M and L dwarfs, respectively.
\label{fig7}}

\clearpage
\plotone{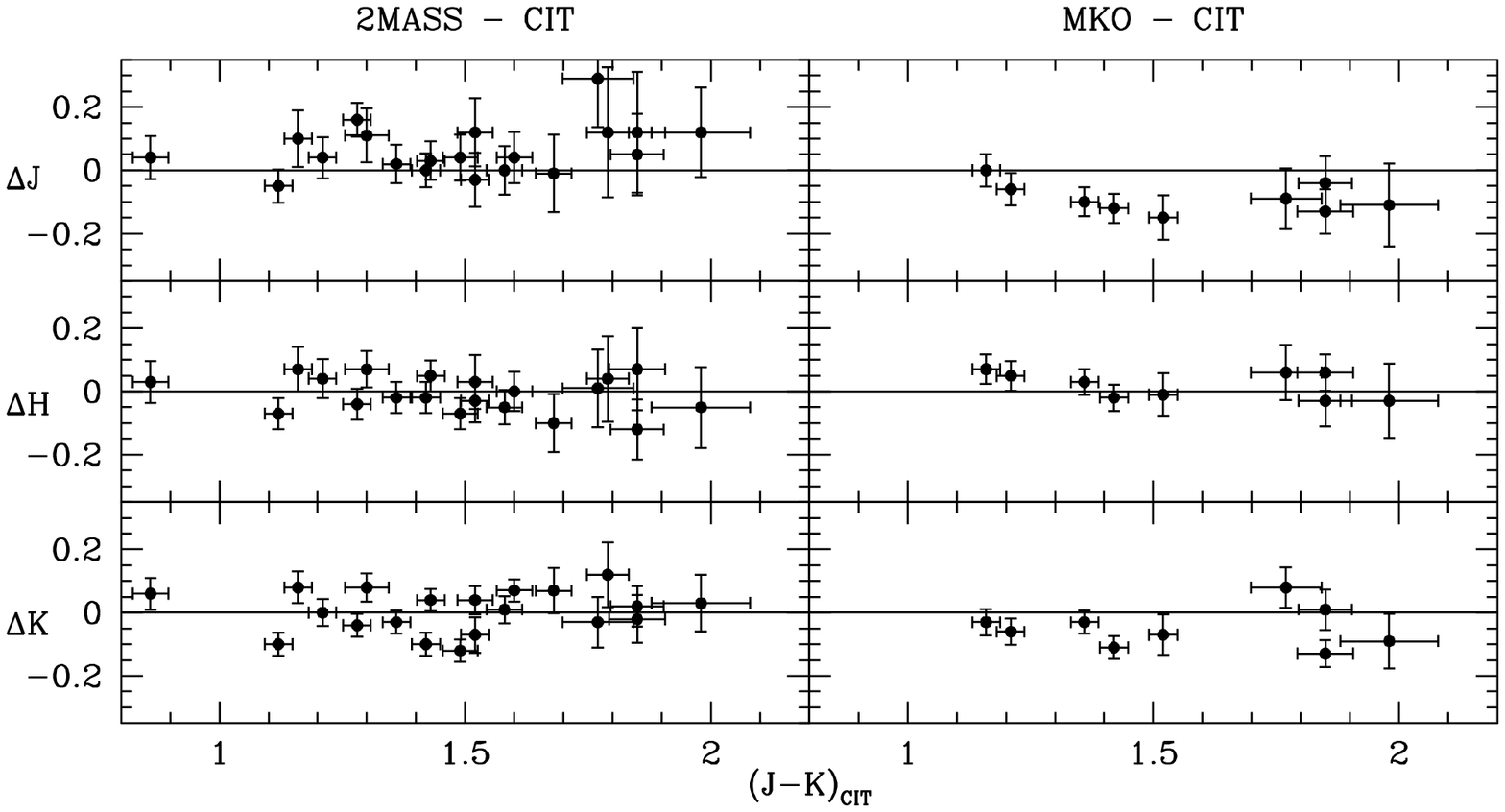}
\clearpage
\plotone{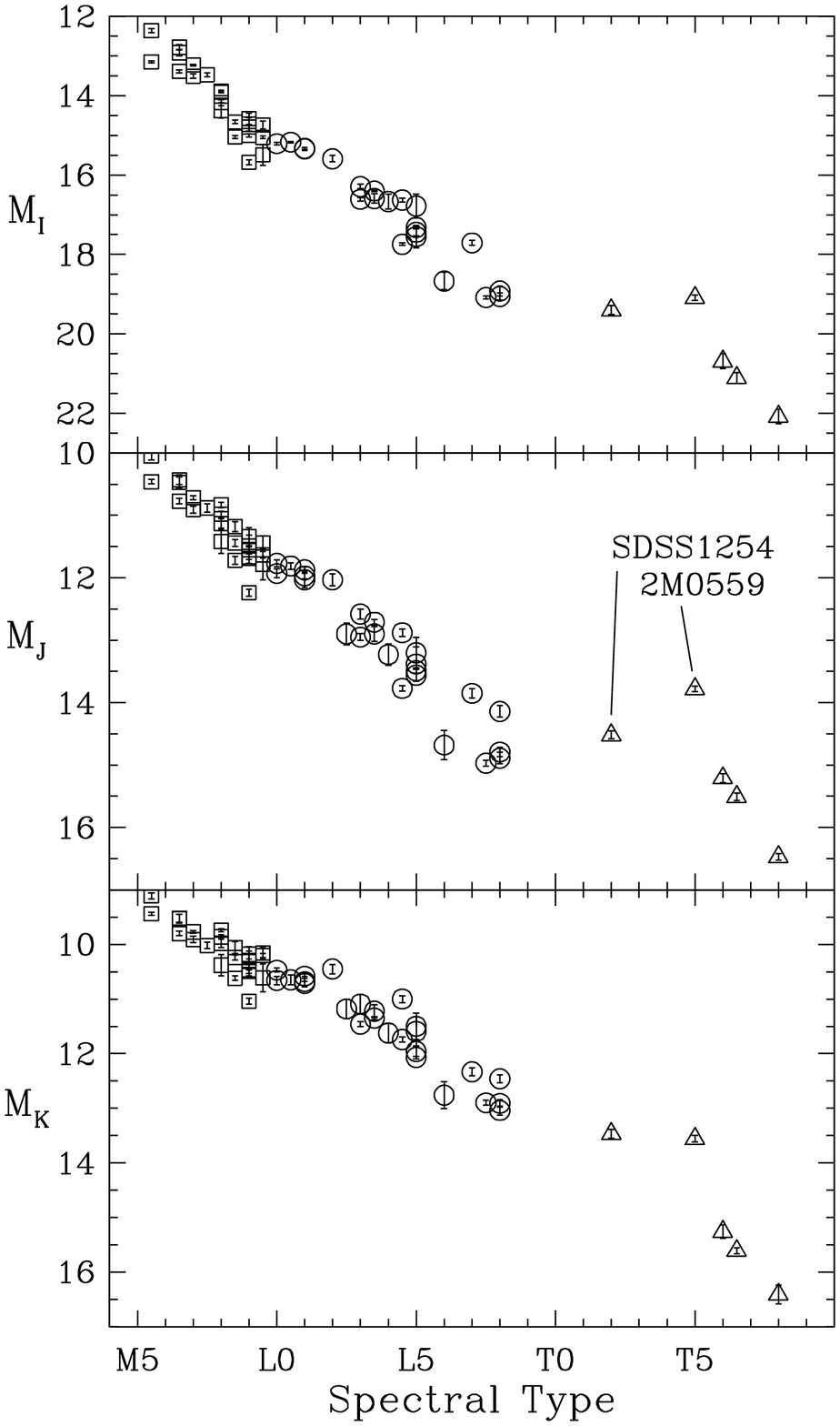}
\clearpage
\plotone{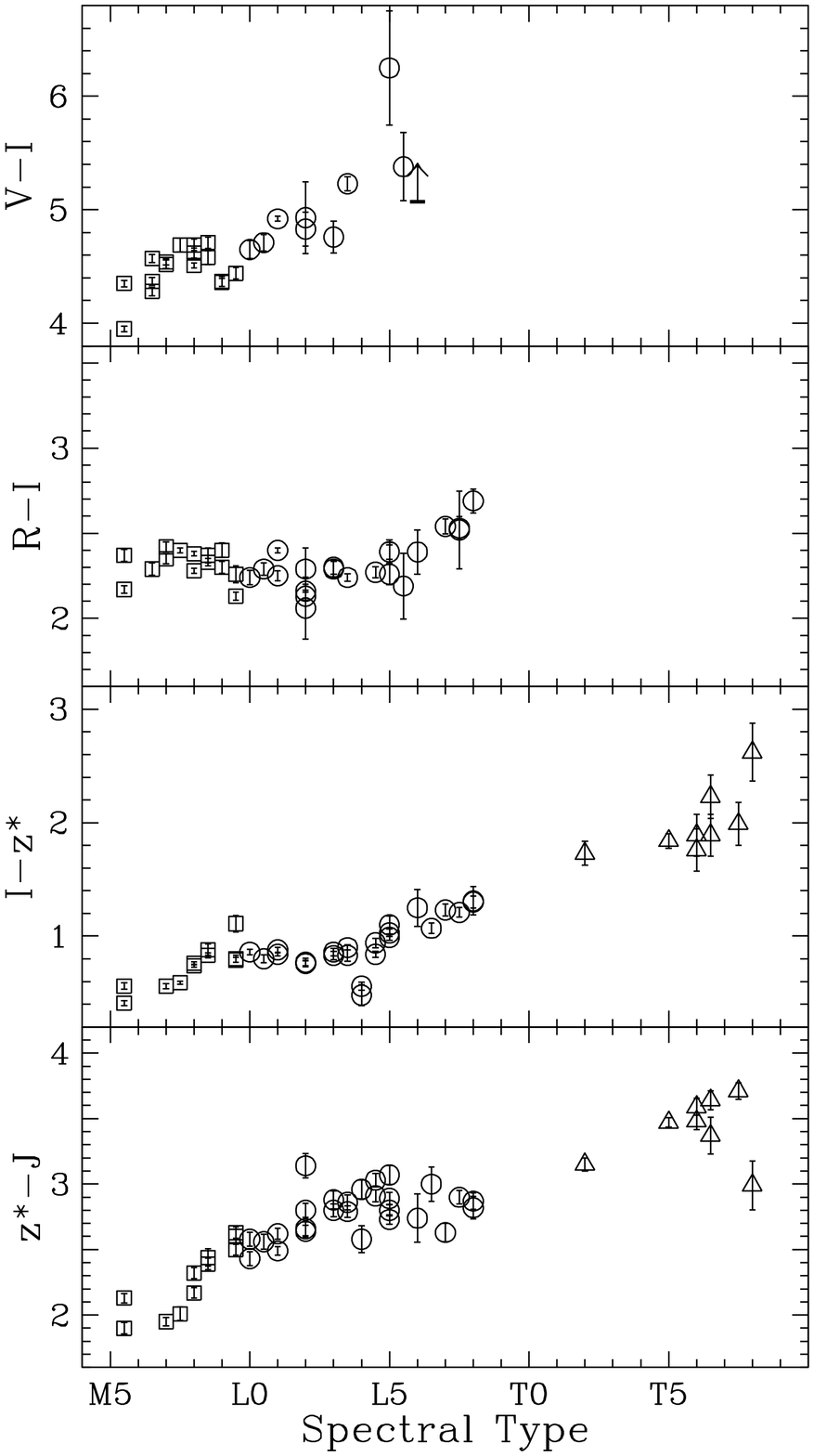}
\clearpage
\plotone{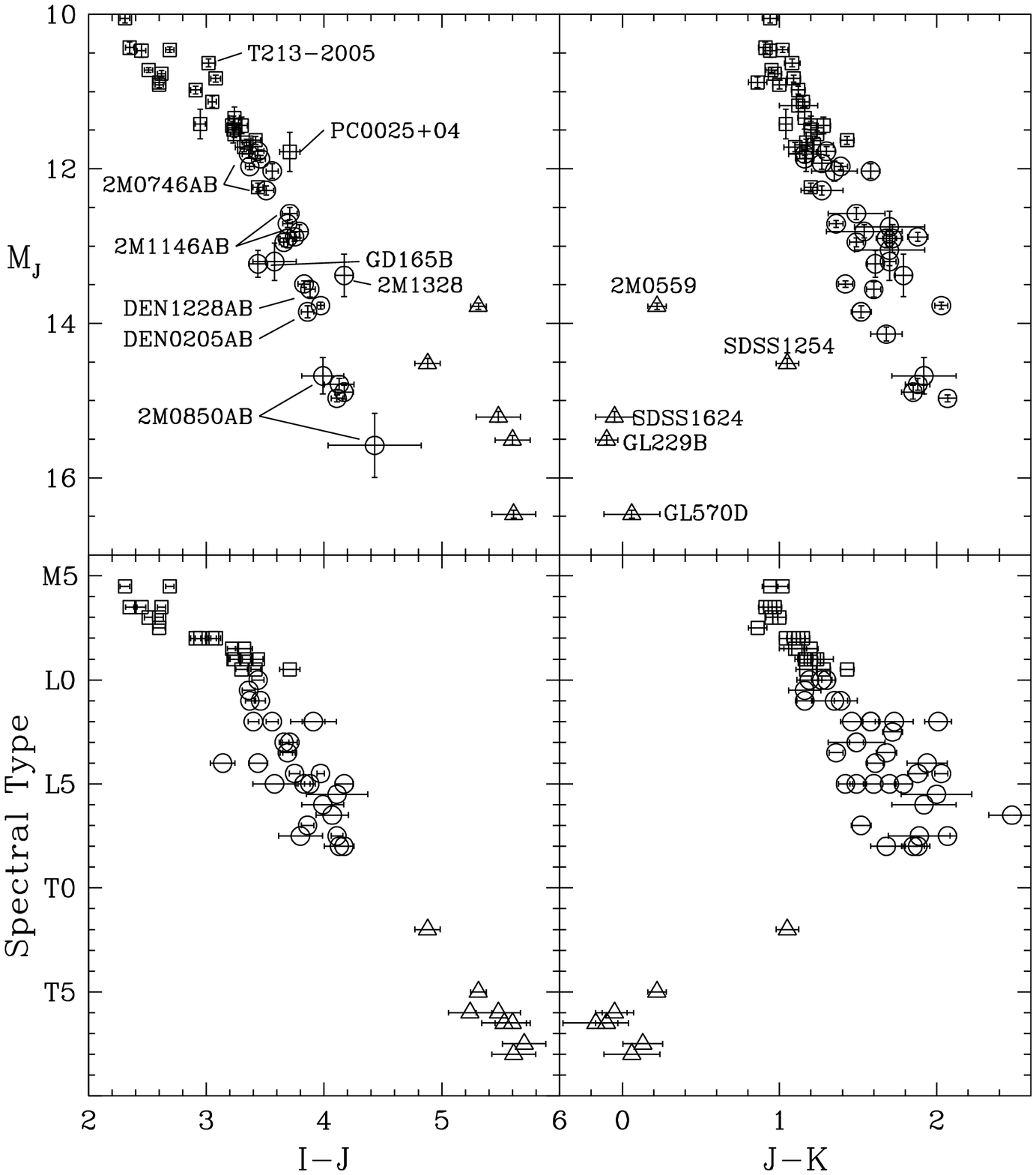}
\clearpage
\plotone{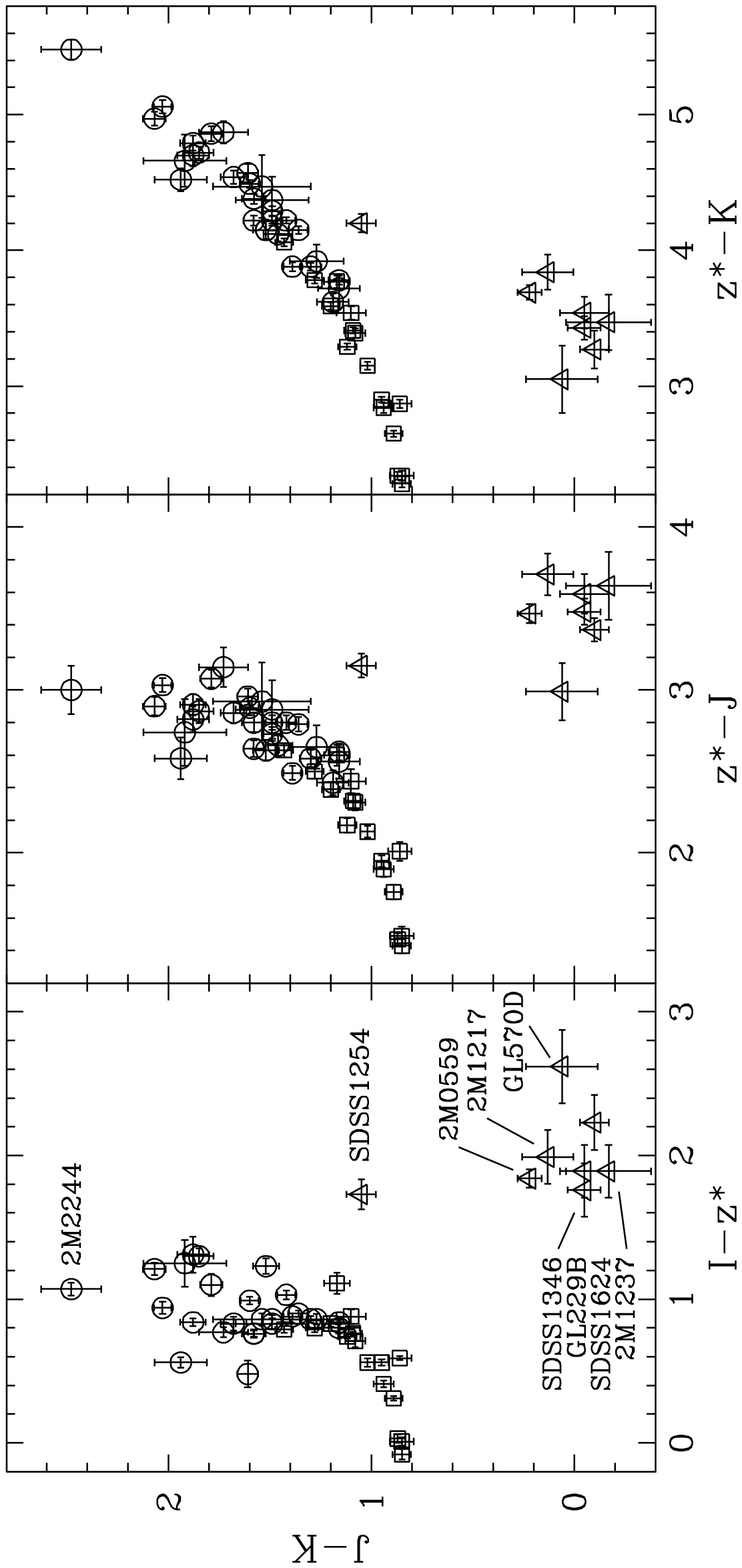}
\clearpage
\plotone{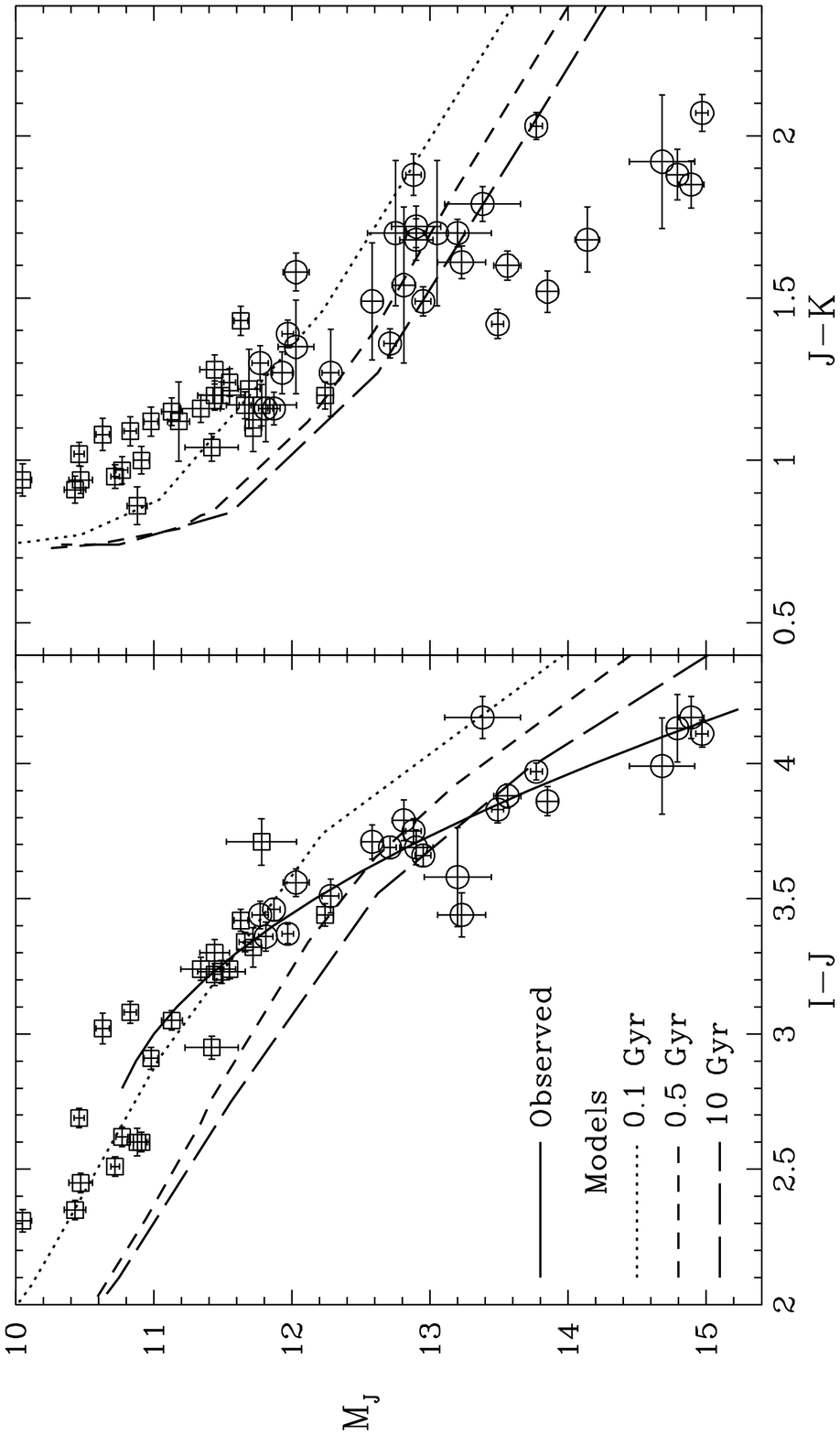}
\clearpage
\plotone{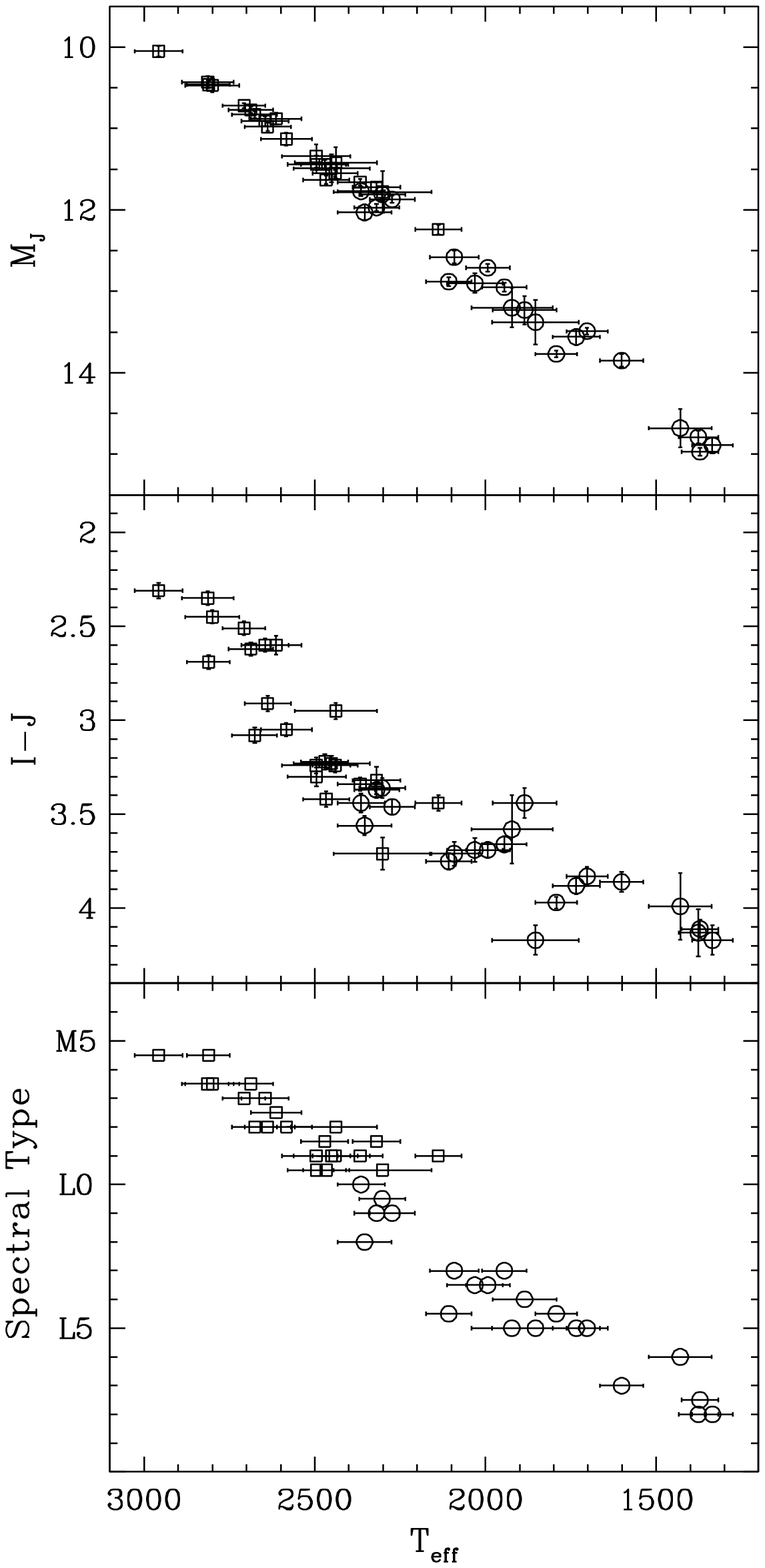}


\clearpage

\begin{thebibliography}{}

\bibitem[Ackerman \& Marley (2001)]{ack01} Ackerman, A., \& Marley, M.
   2001, \apj, 556, 872
\bibitem[Allard et al. (2001)]{all01} Allard, F., Hauschildt, P.H.,
   Alexander, D.R., Tamanai, A., \& Schweitzer, A. 2001, \apj, 556, 357
\bibitem[Bailer-Jones \& Mundt (2001)]{bai01} Bailer-Jones, C.A.L., \&
   Mundt, R. 2001, \aap, 367, 218
% \bibitem[Basri (1998)]{bas98} Basri, G. 1998, in ASP Conf. Ser. 134,
%    Brown Dwarfs and Extrasolar Planets, ed. R. Rebolo, E.L. Mart\'{\i}n,
%    \& M.R. Zapatero Osorio (San Francisco: ASP), 394
\bibitem[Basri (2000)]{basaa00} Basri, G. 2000, ARA\& A, 38, 485
\bibitem[Basri et al. (2000)]{bas00} Basri, G., Mohanty, S., Allard, F.,
   Hauschildt, P.H., Delfosse, X., Mart\'in, E.L., Forveille, T., \&
   Goldman, B. 2000, \apj, 538, 363
\bibitem[Becklin \& Zuckerman (1988)]{bec88} Becklin, E.E., \& Zuckerman, B.
   1988, Nature, 336, 656
\bibitem[Bessell (1990a)]{bes90} Bessell, M.S. 1990a, \pasp, 102, 1181
\bibitem[Bessell (1990b)]{bes90b} Bessell, M.S. 1990b, A\&AS, 83, 357
\bibitem[Burgasser (2001)]{burth} Burgasser, A.J. 2001, Ph.D. Dissertation,
   California Institute of Technology
\bibitem[Burgasser et al. (1999)]{bur99} Burgasser, A.J., et al. 1999,
   \apj, 522, L65
\bibitem[Burgasser et al. (2000a)]{bur00a} Burgasser, A.J., et al. 2000a,
   \aj, 120, 1100
\bibitem[Burgasser et al. (2000b)]{bur00b} Burgasser, A.J., et al. 2000b,
   \apj, 531, L57
\bibitem[Burgasser et al. (2000c)]{bur00c} Burgasser, A.J., et al. 2000c,
   in From Giant Planets to Cool Stars, ASP Conf. Ser. 212,
   ed. C.A. Griffith \& M.S. Marley (San Francisco: ASP), 65
\bibitem[Burgasser et al. (2002)]{bur02} Burgasser, A.J., et al. 2002,
   \apj, 564, 421
\bibitem[Burrows \& Sharp (1999)]{burr99} Burrows, A., \& Sharp, C.M.
   1999, \apj, 512, 843
\bibitem[Burrows et al. (2001)]{burr01} Burrows, A., Hubbard, W.B.,
   Lunine, J.I., \& Liebert, J. 2001, Rev. Mod. Phys., 73, 719
\bibitem[Burrows et al. (2002)]{burr02} Burrows, A., Burgasser, A.J.,
   Kirkpatrick, J.D., Liebert, J., Milsom, J.A., Sudarsky, M.D., \&
   Hubeny, I. 2002, \apj, in press, astro-ph/0109227
\bibitem[Burrows et al. (1997)]{burr97} Burrows, A., Marley, M.,
   Hubbard, W.B., Lunine, J.I., Guillot, T., Saumon, D.,
   Freedman, R., Sudarsky, D., \& Sharp, C. 1997, \apj, 491, 856
\bibitem[Carpenter (2001)]{car01} Carpenter, J.M. 2001, \aj, 121, 2851
%\bibitem[Casali \& Hawarden (1992)]{cas92} Casali, M, \& Hawarden, T. 1992,
%   JCMT-UKIRT Newsl., No.4, 33
\bibitem[Chabrier (2002)]{cha02} Chabrier, G. 2002, \apj, in press,
   astro-ph/0110024
\bibitem[Chabrier \& Baraffe (2000)]{chabar00} Chabrier, G., \& Baraffe, I.
   2000, \araa, 38, 337
\bibitem[Chabrier et al. (2000)]{cha00} Chabrier, G., Baraffe, I.,
   Allard, F., \& Hauschildt, P. 2000, \apj, 542, 464
\bibitem[Clarke et al. (2002)]{cla02} Clarke, F.J., Tinney, C.G., \&
   Covey, K.R. 2002, \mnras, in press, astro-ph/0201162
\bibitem[Dahn et al. (2000)]{dah00} Dahn, C.C., et al. 2000,
   in From Giant Planets to Cool Stars, ASP Conf. Ser. 212,
   ed. C.A. Griffith \& M.S. Marley (San Francisco: ASP), 74
\bibitem[Delfosse et al. (1997)]{del97} Delfosse, X., et al. 1997,
   \aap, 327, L25
\bibitem[Elias et al. (1982)]{eli82} Elias, J.H., Frogel, J.A., Matthews, K.,
   \& Neugebauer, G. 1982, \aj, 87, 1029
\bibitem[Epchtein et al. (1997)]{epc97} Epchtein, N., et al. 1997,
   The Messenger, No. 87, 27
\bibitem[Fan et al. (2000)]{fan00} Fan, X., et al. 2000, \aj, 119, 928
\bibitem[Forrest et al. (1988)]{for88} Forrest, W.J., Skrutskie, M.F.,
   \& Shure, M. 1988, \apj, 330, L119
\bibitem[Geballe et al. (1996)]{geb96} Geballe, T.R., Kulkarni, S.R.,
   Woodward, C.E., \& Sloan, C.G. 1996, \apj, 467, L101
\bibitem[Geballe et al. (2001)]{geb01} Geballe, T.R., Saumon, D.,
   Leggett, S.K., Knapp, G.R., Marley, M.S., \& Lodders, K. 2001,
   \apj, 556, 373
\bibitem[Geballe et al. (2002)]{geb02} Geballe, T.R., et al. 2002,
   \apj, 564, 466
\bibitem[Giclas et al. (1964)]{gic64} Giclas, H. et al. 1964,
   Lowell Obs. Bull., No. 124, v. VI, No. 5
\bibitem[Gilmore et al. (1985)]{gil85} Gilmore, G., Reid, N., \&
   Hewett, P. 1985, \mnras, 213, 257
\bibitem[Gizis et al. (2000)]{giz00} Gizis, J.E., Monet, D.G., Reid, I.N.,
   Kirkpatrick, J.D., Liebert, J., \& Williams, R.J. 2000, \aj, 120, 1085
\bibitem[Gizis et al. (2001)]{giz01} Gizis, J.E., Kirkpatrick, J.D.,
   \& Wilson, J.C. 2001, \aj, 121, 2185
\bibitem[Goldman et al. (1999)]{gol99} Goldman, B., et al. 1999,
   \aap, 351, L5
\bibitem[Golimowski et al. (1998)]{gol98} Golimowski, D.A., Burrows,
   C.J., Kulkarni, S.R., Oppenheimer, B.R., \& Brukardt, B. 1998,
   \aj, 115, 2579
\bibitem[Hauschildt et al. (1997)]{hau97} Hauschildt, P.H., Starrfield, S.,
   Allard, F., \& Alexander, D.R. 1997, \araa, 35, 137
%\bibitem[Hawarden et al. (2001)]{haw01} Hawarden, T.G., Leggett, S.K.,
%   Letawsky, M.B., Ballantyne, D.R., \& Casali, M.M. 2001, \mnras,
%   325, 563
\bibitem[Hawley et al. (1996)]{haw96} Hawley, S.L., Gizis, J.E., \&
   Reid, I.N. 1996, \aj, 112, 2799
\bibitem[The Hipparcos and Tycho Catalogues]{hip97} The Hipparcos and
   Tycho Catalogues 1997, (Noordwijk: ESA)
\bibitem[Irwin et al. (1991)]{irw91} Irwin, M., McMahon, R.G., \&
   Reid, N. 1991, \mnras, 252, 61p
\bibitem[Kirkpatrick (1998)]{kir98} Kirkpatrick, J.D. 1998, in ASP Conf.
   Ser. 134, Brown Dwarfs and Extrasolar Planets, ed. R. Rebolo,
   E.L. Mart\'{\i}n, \& M.R. Zapatero Osorio (San Francisco: ASP), 405
\bibitem[Kirkpatrick et al. (1997a)]{kir97a} Kirkpatrick, J.D.,
   Beichman, C.A., \& Skrutskie, M.F. 1997a, \apj, 476, 311
\bibitem[Kirkpatrick et al. (1997b)]{kir97b} Kirkpatrick, J.D., Henry, T.J.,
   \& Irwin, M.J. 1997b, \aj, 113, 1421
\bibitem[Kirkpatrick et al. (1993)]{kir93} Kirkpatrick, J.D., Henry, T.J.,
   \& Liebert, J. 1993, \apj, 406, 701
\bibitem[Kirkpatrick et al. (1991)]{kir91} Kirkpatrick, J.D., Henry, T.J.,
   \& McCarthy, D.W. 1991, \apjs, 77, 417
\bibitem[Kirkpatrick et al. (1995)]{kir95} Kirkpatrick, J.D., Henry, T.J.,
   \& Simons, D.A. 1995, \aj, 109, 797
\bibitem[Kirkpatrick et al. (1999)]{kir99} Kirkpatrick, J.D., et al.
   1999, \apj, 519, 802 (K99)
\bibitem[Kirkpatrick et al. (2000)]{kir00} Kirkpatrick, J.D., et al.
   2000, \aj, 120, 447 (K00)
\bibitem[Kirkpatrick et al. (2001)]{kir01} Kirkpatrick, J.D., Dahn, C.C.,
   Monet, D.G., Reid, I.N., Gizis, J.E., Liebert, J., \& Burgasser, A.J.
   2001, \aj, 121, 3235
\bibitem[Koerner et al. (1999)]{koe99} Koerner, D.W., Kirkpatrick, J.D.,
   McElwain, M.W., \& Bonaventura, N.R. 1999, \apj, 526, L25
\bibitem[Landolt (1983)]{lan83} Landolt, A.U. 1983, \aj, 88, 439
\bibitem[Landolt (1992)]{lan92} Landolt, A.U. 1992, \aj, 104, 340
\bibitem[Lane et al. (2001)]{lan01} Lane, B.F., Zapatero Osorio, M.R.,
   Britton, M.C., Mart\'{\i}n, E.L., \& Kulkarni, S.R. 2001, \apj, 560, 390
\bibitem[Leggett (1992)]{leg92} Leggett, S.K. 1992, \apjs, 82, 351
\bibitem[Leggett et al. (2000a)]{leg00a} Leggett, S.K., Allard, F.,
   Dahn, C., Hauschildt, P.H., Kerr, T.H., \& Rayner, J. 2000a,
   \apj, 535, 965
\bibitem[Leggett et al. (2001)]{leg01} Leggett, S.K., Allard, F.,
   Geballe, T.R., Hauschildt, P.H., \& Schweitzer, A. 2001, \apj,
   548, 908
\bibitem[Leggett et al. (1998)]{leg98} Leggett, S.K., Allard, F., \&
   Hauschildt, P.H. 1998, \apj, 509, 836
\bibitem[Leggett et al. (1999)]{leg99} Leggett, S.K., Toomey, D.W.,
   Geballe, T.R., \& Brown, R.H. 1999, \apj, 517, L139
\bibitem[Leggett et al. (2000b)]{leg00b} Leggett, S.K., et al. 2000b,
   \apj, 536, L35
\bibitem[Leggett et al. (2002a)]{leg02a} Leggett, S.K., et al. 2002a,
   \apj, 564, 452
\bibitem[Leggett et al. (2002b)]{leg02b} Leggett, S.K., Hauschildt, P.H.,
   Allard, F., Geballe, T.R., \& Baron, E. 2002b, \mnras, in press,
   astro-ph/0112335
\bibitem[Liebert (2000)]{lie00} Liebert, J. 2000,
   in From Giant Planets to Cool Stars, ASP Conf. Ser. 212,
   ed. C.A. Griffith \& M.S. Marley (San Francisco: ASP), 7
\bibitem[Liebert et al.\ (1999)]{lie99} Liebert, J., Kirkpatrick, J.D.,
   Reid, I.N., \& Fisher, M.D. 1999, \apj, 519, 345
\bibitem[Luyten (1979a)]{luy79a} Luyten, W.J. 1979a, NLTT Catalogue,
   Vol. II (Minneapolis: Univ. of Minnesota)
\bibitem[Luyten (1979b)]{luy79b} Luyten, W.J. 1979b, LHS Catalogue,
   2nd Ed. (Minneapolis: Univ. of Minnesota)
\bibitem[Luyten \& Albers (1979)]{lua79} Luyten, W.J. \& Albers, H.
   1979, LHS Atlas (Minneapolis: Univ. of Minnesota)
\bibitem[Marley et al. (2002)]{mar02} Marley, M.S., Seager, S.,
   Saumon, D., Lodders, K., Ackerman, A.S., Freedman, R.S., \& Fan, X.
   2002, \apj, 568, 335
\bibitem[Mart\'{\i}n et al. (1997)]{mar97} Mart\'{\i}n, E.L., Basri, G.,
   Delfosse, X., \& Forveille, T. 1997, \aap, 327, L29
\bibitem[Mart\'in et al. (1999a)]{mar99a} Mart\'in, E.L., Basri, G., \&
   Zapatero Osorio, M.R. 1999a, \aj, 118, 1005
\bibitem[Mart\'{\i}n et al. (1999b)]{mar99b} Mart\'{\i}n, E.L.,
   Brandner, W., \& Basri, G. 1999b, Science, 283, 1718
\bibitem[Mart\'in et al. (1999c)]{mar99c} Mart\'in, E.L., Delfosse, X.,
   Basri, G., Goldman, B., Forveille, T., \& Zapatero Osorio, M.R.
   1999c, \aj, 118, 2466
\bibitem[Mart\'in et al. (2001)]{mar01} Mart\'in, E.L.,
   Zapatero Osorio, M.R., \& Lehto, H.J. 2001, \apj, 557, 822
% \bibitem[Matthews et al. (1996)]{mat96} Matthews, K., Nakjima, T.,
%    Kulkarni, S.R., \& Oppenheimer, B.R. 1996, \aj, 112, 1678
\bibitem[McLean et al. (2001)]{mcl01} McLean, I.S., Prato, L,
   Kim, S.S., Wilcox, M.K., Kirkpatrick, J.D., \& Burgasser, A. 2002,
   \apj, 561, L115
\bibitem[Monet et al. (1992)]{mon92} Monet, D.G., Dahn, C.C.,
   Vrba, F.J., Harris, H.C., Pier, J.R., Luginbuhl, C.B., \&
   Ables, H.D. 1992, \aj, 103, 638
\bibitem[Nakajima et al. (1995)]{nak95} Nakajima, T., Oppenheimer, B.R.,
   Kulkarni, S.R., Golimowski, D.A., Matthews, K., \& Durrance, S.T.
   1995, Nature, 378, 463
\bibitem[Oppenheimer et al. (1995)]{opp95} Oppenheimer, B.R., Kulkarni,
   S.R., Matthews, K., \& Nakajima, T. 1995, Science, 270, 1478
\bibitem[Oppenheimer et al. (1998)]{opp98} Oppenheimer, B.R., Kulkarni,
   S.R., Matthews, K., \& van Kerkwijk, M.H. 1998, \apj, 502, 932
\bibitem[Pavlenko et al. (2000)]{pav00} Pavlenko, Y., Zapatero Osorio,
   M.R., \& Rebolo, R. 2000, \aap, 355, 245
\bibitem[Pier et al. (2000)]{pie00} Pier, J.R., et al. 2000,
   in From Giant Planets to Cool Stars, ASP Conf. Ser. 212,
   ed. C.A. Griffith \& M.S. Marley (San Francisco: ASP), 30
\bibitem[Potter et al. (2002)]{pot02} Potter, D., Mart\'{\i}n, E.L.,
   Cushing, M.C., Baudoz, P., Brandner, W., Guyon, O., \&
   Neuh\"auser, R. 2002, \apj, in press, astro-ph/0201431
% \bibitem[Rebolo et al. (1992)]{reb92} Rebolo, R., Mart\'{\i}n, E.L., \&
%    Magazz\`{u} 1992, \apj, 389, L83
\bibitem[Reid et al. (1999)]{rei99} Reid, I.N., et al. 1999, \apj, 521, 613
\bibitem[Reid et al. (2001a)]{rei01a} Reid, I.N., Burgasser, A.J., Cruz,
   K.L., Kirkpatrick, J.D., \& Gizis, J.E. 2001a, \aj, 121, 1710
\bibitem[Reid et al. (2001b)]{rei01b} Reid, I.N., Gizis, J.E.,
   Kirkpatrick, J.D., \& Koerner, D.W. 2001b, \aj, 121, 489
\bibitem[Reid \& Hawley (2000)]{reihaw00} Reid, I.N., \& Hawley, S.L.
   2000, New Light on Dark Stars (Chichester: Praxis)
\bibitem[Reid et al. (1995)]{rei95} Reid, I.N., Hawley, S.L., \&
   Gizis, J.E. 1995, \aj, 110, 1838
\bibitem[Reid \& Cruz (2002)]{reicru02} Reid, I.N., \& Cruz, K.L.
   2002, \aj, 123, 466
\bibitem[Reid et al. (2002)]{rei02} Reid, I.N., Kirkpatrick, J.D.,
   Liebert, J., Gizis, J.E., Dahn, C.C. \& Monet, D.G. 2002, \aj,
   in press, astro-ph/0204285
\bibitem[Ruiz et al. (1997)]{rui97} Ruiz, M.T., Leggett, S.K., \&
   Allard, F. 1997, \apj, 491, L107
\bibitem[Schneider et al. (1991)]{sch91} Schneider, D.P., Greenstein,
   J.L., Schmidt, M., \& Gunn, J.E. 1991, \aj, 102, 1180
\bibitem[Schneider et al. (2002)]{schn02} Schneider, D.P., et al.
   2002, \aj, 123, 458
% \bibitem[Schultz et al. (1998)]{sch98} Schultz, A.B., et al. 1998,
%    \apj, 492, L181
\bibitem[Schweitzer et al. (2002)]{schw02} Schweitzer, A., Gizis, J.E.,
   Hauschildt, P.H., Allard, F., Howard, E.M., \& Kirkpatrick, J.D. 2002,
   \apj, 566, 435
\bibitem[Schweitzer et al. (2001)]{schw01} Schweitzer, A., Gizis, J.E.,
   Hauschildt, P.H., Allard, F., \& Reid, I.N. 2001, \apj, 555, 368
\bibitem[Skrutskie et al. (1997)]{skr97} Skrutskie, M.F., et al. 1997,
   in The Impact of Large-Scale Near-IR Sky Surveys, ed. F. Garzon et al.
   (Dordrecht: Kluwer), 25
\bibitem[Smith et al. (2002)]{smi02} Smith, J.A., et al. 2002,
   \aj, in press, astro-ph/0201143
\bibitem[Steele \& Howells (2000)]{ste00} Steele, I.A., \& Howells, L.
   2000, \mnras, 313, L43
\bibitem[Stephens et al. (2001)]{ste01} Stephens, D.C., Marley, M.S.,
   Noll, K.S., \& Chanover, N. 2001, \apj, 556, L101
\bibitem[Strauss et al. (1999)]{str99} Strauss, M.A., et al. 1999,
   \apj, 522, L61
\bibitem[Testi et al. (2001)]{tes01} Testi, L., et al. 2001, \apj,
   552, L147
\bibitem[Tinney (1996)]{tin96} Tinney, C.G. 1996, \mnras, 281, 644
\bibitem[Tinney \& Tolley (1999)]{tin99} Tinney, C.G., \& Tolley, A.J.
   1999, \mnras, 304, 119
% \bibitem[Tinney et al. (1997)]{tin97} Tinney, C.G., Delfosse, X., \&
%    Forveille, T. 1997, \apj, 490, L95
\bibitem[Tinney et al. (1993)]{tin93} Tinney, C.G., Mould, J.R., \&
   Reid, I.N. 1993, \aj, 105, 1045
\bibitem[Tinney et al. (1995)]{tin95} Tinney, C.G., Reid, I.N.,
   Gizis, J., \& Mould, J.R. 1995, \aj, 110, 3014
\bibitem[Tinney et al. (1998)]{tin98} Tinney, C.G., Delfosse, X.,
   Forveille, T., \& Allard, F. 1998, \aap, 338, 1066
\bibitem[Tokunaga et al. (2002)]{tok02} Tokunaga, A.T., Simons, D.A.,
   \& Vacca, W.D. 2002, \pasp, 114, 180
\bibitem[Tsvetanov et al. (2000)]{tsv00} Tsvetanov, Z.I., et al. 2000,
   \apj, 531, L61
\bibitem[van Altena et al. (1995)]{van95} van Altena, W.F., Lee,
   J.T., \& Hoffleit, E.D. 1995, The General Catalogue of Trigonometric
   Stellar Parallaxes, Fourth Edition (Schenectady: L. Davis Press)
\bibitem[Wielen (1977)]{wie77} Wielen, R. 1977, \aap, 60, 263
\bibitem[Wilson et al. (2001)]{wil01} Wilson, J.C., Kirkpatrick, J.D.,
   Gizis, J.E., Skrutskie, M.F., Monet, D.G., \& Houck, J.R. 
   2001, \aj, 122, 1989
\bibitem[York et al. (2000)]{yor00} York, D.G., et al. 2000, \aj, 120, 1579

\end{thebibliography}
\end{document}

\begin{deluxetable}{lllll}
\footnotesize
\tablecolumns{5}
\tablewidth{5.5 in}
\pagestyle{empty}
\tablenum{A1}
\tablecaption{Full Names, Coordinates, and Notes}
\tablehead{
\multicolumn{1}{l}{Object} & \multicolumn{2}{c}{Coordinates} &
   \multicolumn{2}{c}{References} \\
\colhead{} & \multicolumn{2}{c}{RA (2000) Dec} & 
\colhead{Coords.} & \colhead{Chart} \\
}
\startdata
LHS102B\tablenotemark{a}  &00:04:33.9 &$-$40:44:06&  1&  1 \\
2MASSW J0015447+351603    &00:15:44.7 &+35:16:03& 2& 2 \\
BRI 0021$-$0214           &00:24:24.6 &$-$01:58:22& 3\tablenotemark{bc}& 3, 4\\
PCC0025+0447              &00:27:41.4 &+05:03:50& 5\tablenotemark{b}&  5 \\
2MASSW J0036159+182110    &00:36:15.9 &+18:21:10&  2&  2 \\
CTI 012657.5+280202       &01:27:39.2 &+28:05:55& 4\tablenotemark{b}& 4 \\  
2MASSW J0149090+295613    &01:49:09.0 &+29:56:13&      6&      6 \\
DENIS$-$P J0205.4$-$1159\tablenotemark{d} &02:05:29.0 &$-$11:59:25& 7& 7 \\  
2MASSI J0235599$-$233120\tablenotemark{e} &02:35:59.9 &$-$23:31:20&  8& 99 \\
LP771$-$21 = BRI 0246$-$1703 &02:48:41.0 &$-$16:51:19& 9\tablenotemark{f}& 9\\
TVLM832$-$10443           &02:52:26.4 &+00:56:23& 10\tablenotemark{g}& 10 \\ 
LP412$-$31                &03:20:58   &+18:54.6 & 11\tablenotemark{b}& 4 \\  
2MASSW J0326137+295015    &03:26:13.7 &+29:50:15&     12&     12 \\
SDSSp J033035.13$-$002534.5  &03:30:35.1 &$-$00:25:35&   13&   13 \\
LP944$-$20 = BRI 0337$-$3535 &03:39:34.6 &$-$35:25:51& 9\tablenotemark{h}& 9\\
2MASSP J0345432+254023    &03:45:43.2 &+25:40:23&     12&     12 \\
SDSSp J041320.38$-$011424.9  &04:13:20.4 &$-$01:14:25&   13&   13 \\
LHS191 = LP535$-$12       &04:26:21   &+03:36.5 &   14&   15 \\
SDSSp J053951.99$-$005902.0  &05:39:52.0 &$-$00:59:02&   13&   13 \\
2MASSI J0559191$-$140448  &05:59:19.1 &$-$14:04:48&   16&     16 \\
Gl229B                    &06:10:34.7 &$-$21:51:46& 17\tablenotemark{i}& 18 \\
2MASSI J0746425+200032    &07:46:42.5 &+20:00:32&      2&      2 \\
2MASSI J0825196+211552    &08:25:19.6 &+21:15:52&      2&      2 \\
2MASSs J0850359+105716    &08:50:35.9 &+10:57:16&     12&     12 \\
LHS2065 = LP666$-$9       &08:53:36   &$-$03:29.5 &   14&   15 \\
2MASSW J0912145+145940\tablenotemark{j} &09:12:14.5 &+14:59:40& 19& 19 \\
TVLM213$-$2005            &10:21:27.8 &+50:55:03& 10\tablenotemark{g}& 10 \\
2MASSI J1022148+411426\tablenotemark{k} &10:22:14.8 &+41:14:26& 19& 19 \\
DENIS$-$P J1058.7$-$1548\tablenotemark{d} &10:58:46.5 &$-$15:48:00& 7& 7 \\  
2MASSW J1112257+354813\tablenotemark{l} &11:12:25.7 &+35:48:13&  2&  2,20 \\
LHS2397a = LP732$-$94     &11:21:49.7 &$-$13:13:07& 9\tablenotemark{m}& 9 \\
2MASSW J1146345+223053    &11:46:34.5 &+22:30:53&     12&     12 \\
LHS2471 = LP553$-$59      &11:53:53   &+07:00.3 &   14&   15 \\
2MASSW J1217111$-$031113  &12:17:11.1 &$-$03:11:13&   21&   21 \\
BRI 1222$-$1222           &12:24:52.4 &$-$12:38:31& 9\tablenotemark{n}& 9 \\
DENIS$-$P J1228.2$-$1547\tablenotemark{d} &12:28:13.8 &$-$15:47:11& 7& 7 \\  
2MASSW J1237392+652615    &12:37:39.2 &+65:26:15&   21&   21 \\
SDSS J125453.90$-$012247.4 &12:54:53.9 &$-$01:22:47&   22&     22 \\
Kelu-1                    &13:05:40.2 &$-$25:41:06& 23\tablenotemark{o}& 23 \\
2MASSW J1328550+211449    &13:28:55.0 &+21:14:49&     12&     12 \\
SDSSp J134646.45$-$003150.4  &13:46:46.5 &$-$00:31:50&   24&   24 \\
LHS2828 = LP499$-$5       &13:57:02   &+08:30.1 &   14&   15 \\
GD165B                    &14:24:39.3 &+09:17:15& 25\tablenotemark{bp}& 4 \\
LHS2924 = LP271-25        &14:28:42   &+33:10.6 &   14&   15 \\
LHS2930 = LP98-79         &14:30:40   &+59:43.2 &   14&   15 \\
2MASSW J1439284+192915    &14:39:28.4 &+19:29:15&   12&   12 \\
HD130948BC\tablenotemark{q} &14:50:15.8 &+23:54:43& 17&   26 \\
Gl569B\tablenotemark{r}   &14:54:29.3 &+16:06:04&   17&   27 \\
LHS3003 = LP914-54        &14:56:38   &$-$28:09.8&  14&   15 \\
2MASSW J1457150$-$212148\tablenotemark{s} &14:57:15.0 &$-$21:21:48& 28& 28 \\
LHS3005 = G136$-$39       &14:57:20   &+14:59.0 &   14&   15 \\
TVLM513$-$46546           &15:01:08.2 &+22:50:02& 10\tablenotemark{g}& 10 \\  
2MASSW J1507476$-$162738  &15:07:47.6 &$-$16:27:38&    2&      2 \\
TVLM868$-$110639          &15:10:17.1 &$-$02:36:07& 10\tablenotemark{b}& 10 \\
2MASSW J1523226+301456\tablenotemark{t} &15:23:22.6 &+30:14:56&  2&  2,20 \\
2MASSW J1553214+210907    &15:53:21.4 &+21:09:07&   12&  12 \\
2MASSW J1620261$-$041631\tablenotemark{u} &16:20:26.1 &$-$04:16:31& 19& 19 \\
SDSS J162414.37+002915.6  &16:24:14.4 &+00:29:16&     29&     29 \\
2MASSW J1632291+190441    &16:32:29.1 &+19:04:41&     12&     12 \\
LHS429 = vB8              &16:55:34   &$-$08:23.6 &   14&   15 \\
2MASSW J1658038+702702    &16:58:03.8 &+70:27:02&     99&     99 \\
L1421$-$36 = G170$-$28 = LP332$-$9 &17:14:42   &+26:56.0 &   11&   30 \\
2MASSI J1726000+153819    &17:26:00.0 &+15:38:19&    2&    2 \\
LHS3339 = LP102$-$330     &17:55:33   &+58:24.4 &   14&   15 \\
LHS3350 = G182$-$34       &18:01:16   &+35:35.9 &   14&   15 \\
2MASSW J1841086+311727    &18:41:08.6 &+31:17:27&   12&  12 \\
LHS3406 = LP229$-$30      &18:43:22   &+40:40.4 & 14\tablenotemark{v}& 15 \\
\tablebreak
2MASSW J2101154+175658    &21:01:15.4 &+17:56:58&   12&  12 \\
2MASSW J2208136+292121    &22:08:13.6 &+29:21:21&   12&  12 \\
GRH 2208$-$2007&22:10:50.0&$-$19:52:13&31\tablenotemark{t}&9\tablenotemark{w}\\
2MASSW J2224438$-$015852  &22:24:43.8 &$-$01:58:52&    2&      2 \\
2MASSW J2244316+204343    &22:44:31.6 &+20:43:43&   99&   99 \\

\tablenotetext{a}{Faint companion to LHS102 = LP988$-$102 = GJ1001.}
\tablenotetext{b}{Coordinates precessed from 1950.0 values given in Ref.
  to equinox 2000.0 (but {\it not} epoch 2000.0).}
\tablenotetext{c}{Coordinates for Epoch 1989.8.  See remarks in Ref.4 regarding
  confusion with earlier discovery of this object by Luyten.}
\tablenotetext{d}{In original discovery paper the name employed used ``DBD"
  rather than ``DENIS$-$P."  However, the "DENIS$-$P" became standard
  nomenclature for such objects discovered in the "Minisurvey" from DENIS
  (see Tinney et al. 1998).  In the 1997 and 1998 papers the names included
  the RA portion only to integer minutes of time; in various later papers
  (e.g., K99; K00), the name includes decimal minutes of time.}
\tablenotetext{e}{Faint companion to CD$-$24\arcdeg1154 = HD16270 = GJ1048.}
\tablenotetext{f}{Coordinates for Epoch 1988.9.}
\tablenotetext{g}{2000.0 coordinates from SIMBAD.}
\tablenotetext{h}{Coordinates for Epoch 1976.9.}
\tablenotetext{i}{Faint companion to BD$-$21\arcdeg1377 = HD42581 = LHS1827 =
  Gliese 229.  Coordinates are for Gl229A at Epoch 1991.25.  The separation
  and offset to Gl229B is 7.7 arcsec at 163.7 degrees (Ref.18).}
\tablenotetext{j}{Faint companion to BD+15\arcdeg2003AB = HD79096AB =
  LHS2114 = Gliese 337AB.}
\tablenotetext{k}{Faint companion to HR4067 = BD+41\arcdeg2076 = HD89744.}
\tablenotetext{l}{Faint companion to BD+36\arcdeg2162 = HD97334 = Gliese 417.}
\tablenotetext{m}{Coordinates for Epoch 1985.4.}
\tablenotetext{n}{Coordinates for Epoch 1984.3.}
\tablenotetext{o}{Coordinates for Epoch 1993.2.}
\tablenotetext{p}{Coordinates for GD165A at Epoch 1988.7.
  GD165B is located 0.7"W and 4.2"S of GD165A (Ref.25).}
\tablenotetext{q}{Pair of faint companions to HD130948 = BD+24\arcdeg2786
  = HR5534 = Gliese 564.  Coordinates are for HD130948A.  The separation
  and offset to HD130948BC is 3 arcsec at 110 degrees (Ref.26).}
\tablenotetext{r}{Pair of faint companions to G138$-$28 = BD+16\arcdeg2708
  = Gliese 569.  Coordinates are for Gl569A.  The separation and offset to
  Gl569B is 5 arcsec at 18 degrees (Ref.27).}
\tablenotetext{s}{Faint companion to the triple system composed of
  BD$-$20\arcdeg4125 = HD131977, and BD$-$20\arcdeg4123AB = HD131976AB,
  or Gliese 570ABC.}
\tablenotetext{t}{Faint companion to BD+30\arcdeg2653AB = HD137107/8 =
  Gliese 584AB.}
\tablenotetext{u}{Faint companion to Gliese 618.1.}
\tablenotetext{v}{The 2000 RA coordinate given in LHS Catalogue (18:42:22)
  seems to be incorrect by approximately 1.0 min.  The more correct value is
  quoted here.}
\tablenotetext{w}{Coordinates for Epoch 1976.0.  As noted in Ref.9,
  the chart given in Ref.31 is incorrect.}

\tablerefs{ \\
1.  Goldman et al. (1999) \\
2.  K00 \\
3.  Irwin et al. (1991) \\
4.  Kirkpatrick et. al. (1995) \\
5.  Schneider et al. (1991) \\
6.  Liebert et al. (1999) \\
7.  Delfosse et al. (1997) \\
8.  Gizis et al. (2001) \\
9.  Tinney (1996) \\
10. Tinney et al. (1995) \\
11. Luyten (1979a - NLTT) \\
12. K99 \\
13. Fan et al. (2000) \\
14. Luyten (1979b - LHS) \\
15. Luyten \& Albers (1979 - LHS Atlas) \\
16. Burgasser et al. (2000a) \\
17. The Hipparcos and Tycho Catalogues, Vol.6, (1997) \\
18. Golimowski et al. (1998) \\
19. Wilson et al. (2001) \\
20. Kirkpatrick et al. (2001) \\
21. Burgasser et al. (1999) \\
22. Leggett et al. (2000b) \\
23. Ruiz et al. (1997) \\
24. Tsvetanov et al. (2000) \\
25. Becklin \& Zuckerman (1988) \\
26. Potter et al. (2002) \\
27. Forrest et al. (1988) \\
28. Burgasser et al. (2000b) \\
29. Strauss et al. (1999) \\
30. Giclas et al. (1964) \\
31. Gilmore et al. (1985) \\
99. Unpublished \\
}
\enddata
\end{deluxetable}